\PassOptionsToPackage{unicode}{hyperref}
\PassOptionsToPackage{hyphens}{url}
\PassOptionsToPackage{dvipsnames,svgnames,x11names}{xcolor}
\documentclass[
12pt]{article}
\usepackage{setspace}

\usepackage{siunitx}   
\usepackage{xr} 
\usepackage{amsmath,amssymb}
\usepackage{mathtools}
\usepackage{enumitem}
\usepackage{soul}
\usepackage{xcolor}
\setulcolor{red}   
\setstcolor{red}   
\usepackage{amsthm}
\usepackage{multirow}
\usepackage{iftex}
\ifPDFTeX
\usepackage[T1]{fontenc}
\usepackage[utf8]{inputenc}
\usepackage{textcomp} 
\else 
\usepackage{unicode-math}
\defaultfontfeatures{Scale=MatchLowercase}
\defaultfontfeatures[\rmfamily]{Ligatures=TeX,Scale=1}
\fi
\usepackage{lmodern}
\ifPDFTeX\else  
\fi
\IfFileExists{upquote.sty}{\usepackage{upquote}}{}
\IfFileExists{microtype.sty}{
	\usepackage[]{microtype}
	\UseMicrotypeSet[protrusion]{basicmath} 
}{}
\makeatletter
\@ifundefined{KOMAClassName}{
	\IfFileExists{parskip.sty}{%
		\usepackage{parskip}
	}{
		\setlength{\parindent}{0pt}
		\setlength{\parskip}{6pt plus 2pt minus 1pt}}
}{
	\KOMAoptions{parskip=half}}
\makeatother
\usepackage{xcolor}
\makeatletter
\ifx\paragraph\undefined\else
\let\oldparagraph\paragraph
\renewcommand{\paragraph}{
	\@ifstar
	\xxxParagraphStar
	\xxxParagraphNoStar
}
\newcommand{\xxxParagraphStar}[1]{\oldparagraph*{#1}\mbox{}}
\newcommand{\xxxParagraphNoStar}[1]{\oldparagraph{#1}\mbox{}}
\fi
\ifx\subparagraph\undefined\else
\let\oldsubparagraph\subparagraph
\renewcommand{\subparagraph}{
	\@ifstar
	\xxxSubParagraphStar
	\xxxSubParagraphNoStar
}
\newcommand{\xxxSubParagraphStar}[1]{\oldsubparagraph*{#1}\mbox{}}
\newcommand{\xxxSubParagraphNoStar}[1]{\oldsubparagraph{#1}\mbox{}}
\fi
\makeatother

\usepackage{longtable,booktabs,array}
\usepackage{calc} 
\usepackage{etoolbox}
\makeatletter
\patchcmd\longtable{\par}{\if@noskipsec\mbox{}\fi\par}{}{}
\makeatother
\IfFileExists{footnotehyper.sty}{\usepackage{footnotehyper}}{\usepackage{footnote}}
\makesavenoteenv{longtable}
\usepackage{graphicx}
\makeatletter
\def\maxwidth{\ifdim\Gin@nat@width>\linewidth\linewidth\else\Gin@nat@width\fi}
\def\maxheight{\ifdim\Gin@nat@height>\textheight\textheight\else\Gin@nat@height\fi}
\makeatother
\setkeys{Gin}{width=\maxwidth,height=\maxheight,keepaspectratio}
\makeatletter
\def\fps@figure{htbp}
\makeatother

\makeatletter
\@ifpackageloaded{caption}{}{\usepackage{caption}}
\AtBeginDocument{%
	\ifdefined\contentsname
	\renewcommand*\contentsname{Table of contents}
	\else
	\newcommand\contentsname{Table of contents}
	\fi
	\ifdefined\listfigurename
	\renewcommand*\listfigurename{List of Figures}
	\else
	\newcommand\listfigurename{List of Figures}
	\fi
	\ifdefined\listtablename
	\renewcommand*\listtablename{List of Tables}
	\else
	\newcommand\listtablename{List of Tables}
	\fi
	\ifdefined\figurename
	\renewcommand*\figurename{Figure}
	\else
	\newcommand\figurename{Figure}
	\fi
	\ifdefined\tablename
	\renewcommand*\tablename{Table}
	\else
	\newcommand\tablename{Table}
	\fi
}
\@ifpackageloaded{float}{}{\usepackage{float}}
\floatstyle{ruled}
\@ifundefined{c@chapter}{\newfloat{codelisting}{h}{lop}}{\newfloat{codelisting}{h}{lop}[chapter]}
\floatname{codelisting}{Listing}

\makeatother
\makeatletter
\@ifpackageloaded{caption}{}{\usepackage{caption}}
\@ifpackageloaded{subcaption}{}{\usepackage{subcaption}}
\makeatother

\ifLuaTeX
\usepackage{selnolig}  
\fi
\usepackage[sort]{natbib}
\usepackage{bookmark}

\IfFileExists{xurl.sty}{\usepackage{xurl}}{} 
\hypersetup{
	pdftitle={Title},
	pdfauthor={Author 1; Author 2},
	pdfkeywords={3 to 6 keywords, that do not appear in the title},
	colorlinks=true,
	linkcolor={blue},
	filecolor={blue},
	citecolor={Blue},
	urlcolor={Blue},
	pdfcreator={LaTeX via pandoc}}

\global\long\def\R{\mathbb{R}}%
\global\long\def\N{\mathbb{N}}%
\global\long\def\tod{\overset{d}{\to}}%
\global\long\def\e{\mathbb{E}}%
\global\long\def\f{\mathcal{F}}%
\global\long\def\diag{\text{diag}}%
\global\long\def\var{\mathrm{Var}}%
\global\long\def\bs#1{\boldsymbol{#1}}%
\global\long\def\trans{\top}%
\global\long\def\vec{\mathrm{vec}}%

\numberwithin{equation}{section}
\theoremstyle{plain}

\theoremstyle{definition}

\theoremstyle{plain}
\newtheorem{thm}{\protect\theoremname}
\theoremstyle{remark}
\newtheorem{rem}{\protect\remarkname}
\theoremstyle{plain}

\theoremstyle{plain}
\newtheorem{proposition}{\protect\propositionname}

\allowdisplaybreaks[4]
\usepackage{babel}
\usepackage{appendix}\usepackage{titlesec}

\usepackage{tabularray}
\usepackage{verbatim}
\theoremstyle{plain}
\newtheorem{assumption}{\protect\assumptionname}

\theoremstyle{definition}
\newtheorem{myalgorithm}{\protect\algorithmname}

\renewcommand{\hat}{\widehat}

\providecommand{\definitionname}{Definition}

\providecommand{\lemmaname}{Lemma}

\providecommand{\propositionname}{Proposition}
\providecommand{\remarkname}{Remark}

\providecommand{\theoremname}{Theorem}

\makeatother
\providecommand{\algorithmname}{Algorithm}
\providecommand{\assumptionname}{Assumption}

\providecommand{\definitionname}{Definition}
\providecommand{\lemmaname}{Lemma}
\providecommand{\remarkname}{Remark}
\providecommand{\theoremname}{Theorem}

\global\long\def\R{\mathbb{R}}%
\global\long\def\N{\mathbb{N}}%
\global\long\def\tod{\overset{d}{\to}}%
\global\long\def\e{\mathbb{E}}%
\global\long\def\f{\mathcal{F}}%
\global\long\def\diag{\text{diag}}%
\global\long\def\var{\mathrm{Var}}%
\global\long\def\bs#1{\boldsymbol{#1}}%
\global\long\def\trans{\top}%
\global\long\def\vec{\mathrm{vec}}%

\def\boxit#1{\vbox{\hrule\hbox{\vrule\kern6pt
			\vbox{\kern6pt#1\kern6pt}\kern6pt\vrule}\hrule}}

\begin{document}
\def\spacingset#1{\renewcommand{\baselinestretch}%
	{#1}\small\normalsize} \spacingset{1}


		\title{\bf A New and Efficient Debiased Estimation of General Treatment Models by Balanced Neural Networks Weighting\thanks{Address for correspondence: Zheng Zhang, Center for Applied Statistics, Institute of Statistics and Big Data, Renmin University of China, Room 717, Chongde Building (West), Haidian District, Beijing, China.  Email: \href{Email:zhengzhang@ruc.edu.cn}{zhengzhang@ruc.edu.cn}. Running head: Balanced Neural Networks Weighting. Zeqi Wu and Meilin Wang are the co-first authors.}}
		\author{Zeqi Wu\thanks{Center for Applied Statistics, Institute of Statistics and Big Data, Renmin University of China, Beijing, China.} \and Meilin Wang\footnotemark[2] \and Wei Huang\thanks{School of Mathematics and Statistics, University of Melbourne, Melbourne, Australia.}
            \and Zheng Zhang\footnotemark[2]
		}
		\date{}
		\maketitle

\bigskip
\begin{abstract}
Estimation and inference of treatment effects under unconfounded treatment assignments often suffer from bias and the `curse of dimensionality' due to the nonparametric estimation of nuisance parameters for high-dimensional confounders. Although debiased state-of-the-art methods have been proposed for binary treatments under particular treatment models, they can be unstable for small sample sizes. Moreover, directly extending them to general treatment models can lead to computational complexity. We propose a balanced neural networks weighting method for general treatment models, which leverages deep neural networks to alleviate the curse of dimensionality while retaining optimal covariate balance
through calibration, thereby achieving debiased and robust estimation. Our method accommodates a wide range of treatment models, including average, quantile, distributional, and asymmetric least squares treatment effects, for discrete, continuous, and mixed treatments. Under regularity conditions, we show that our estimator achieves rate double robustness and $\sqrt{N}$-asymptotic normality, and its asymptotic variance achieves the semiparametric efficiency bound. We further develop a statistical inference procedure based on weighted bootstrap, which avoids estimating the efficient
influence/score functions. Simulation results reveal that the proposed method consistently outperforms existing alternatives, especially when the sample size is small. Applications to the 401(k) dataset and the Mother's Significant Features dataset further illustrate the practical value of the method for estimating both average and quantile treatment effects under binary and continuous treatments, respectively. 
\end{abstract}
\noindent%
\vfill

\newpage
\doublespacing

\section{Introduction}

Causal inference based on observational data is popular in statistics
and other sciences. The central condition for the identification
is the unconfoundedness that requires the treatment to be conditionally independent of the potential outcomes given observed
confounders \citep{rosenbaum1983central}. Based on such a condition
with a moderate dimension of confounders, extensive semi-parametric
inference methods have been developed in the past decades, see \citet{hirano2003Efficient,chen2008,firpo2007Efficient,ai2021Unified,chan2016Globally}.
However, researchers believe that the unconfoundedness assumption is more plausible to hold with many confounders/covariates (\citealp{farrell2015Robust,fan2022Estimation,athey2018Approximate,su2019Nonseparable}).
In such a case, those methods suffer from the
so-called ``curse of dimensionality'' due to the conventional nonparametric estimation of the nuisance function(s).

To address the curse of dimensionality, modern machine learning techniques have been applied to estimate causal effects, with most existing work focusing on binary treatments \citep[e.g.,][]{chernozhukov2018Double, kallus2022Localized}. While machine learning methods offer flexibility and robustness, they often introduce a non-negligible asymptotic bias. To enable valid inference, double/debiased machine learning (DML) methods have been developed. In particular, \citet{chernozhukov2018Double} first proposed estimating the average treatment effects by solving a score equation based on its efficient influence function, where the nuisance parameters -- the propensity score and conditional mean responses -- are estimated via machine learning. The key driver of DML is that the efficient influence function is orthogonal to these nuisance functions, effectively mitigating the bias introduced by machine learning estimation. Later, \citet{kallus2022Localized} extended the DML idea to estimate the quantile treatment effect for binary treatments.

For continuous treatment effect estimators, asymptotic bias and the curse of dimensionality can be alleviated if the estimator is rate doubly robust -- that is, the estimation error depends on the product of the convergence rates of nuisance estimators, which is substantially smaller than each rate individually. Specifically, \citet{kennedy2017Nonparametric}
proposed a nonparametric estimator for the average dose-response function 
using an augmented
pseudo-outcome that is conditionally orthogonal
to the nuisance: the marginal density of the treatment, the generalized propensity score, and the conditional mean response
function. This yields a rate-doubly robust property, and they estimate the nuisances using machine learning. Adapting this idea to multi-valued treatments, \citet{colangelo2023double} estimated the augmented pseudo-outcome by applying kernel smoothing to the marginal treatment density and employing machine learning methods for the generalized propensity score and the conditional mean response function.
Under some high-level conditions, their estimator is asymptotically normal and retains rate-double robustness.

In summary, existing DML approaches target specific treatment effect models -- binary average/quantile treatment effects or average dose-response function -- and rely on estimating the efficient influence function, which involves the inverse of the propensity score. Estimating propensity score is then particularly challenging with high-dimensional covariates, as it can easily lead to instability due to the inversion \citep{kang2007demystifying,kallus2022Localized}. Moreover, the efficient influence functions for many causal parameters are complicated, e.g., involving the inverse probability density of potential outcomes in binary quantile treatment effect \citep{firpo2007Efficient}, and the projection of the nuisance functions onto treatment and covariate spaces, making estimation and inference computationally complex.

This paper considers a general treatment model that identifies the causal effect as a weighted (possibly non-smooth) optimization, where the weighting
function -- called the \textit{stabilized weighting function} --  is defined as the ratio of the marginal probability function of the treatment to the generalized propensity score. We propose a new debiased approach based on a balanced calibration of an initial estimate of the stabilized weighting function, bypassing the need to estimate the efficient influence function. The initial estimate is obtained by minimizing the squared loss of the stabilized weighting function, thereby alleviating the instability caused by taking the inverse of the generalized propensity score. To tackle the dimensionality problem, we use deep neural networks to estimate the stabilized weighting function, and apply machine learning methods (e.g., deep neural networks, random forests, etc.) to estimate the conditional response function involved in the debiased calibration. We refer to this method as balanced neural networks weighting (BNNW). 

Under some regularity conditions, our novel BNNW estimator of the treatment effect parameter achieves the rate-double robustness, $\sqrt{N}$-asymptotic normality, and semiparametric efficiency.
Our results apply to a wide range of treatment models, including average, quantile, distributional, and asymmetric least squares treatment effects, for discrete, continuous and mixed treatments. While similar general treatment models have been considered in the literature \citep[e.g.,][]{ai2021Unified}, their methods lack rate-double robustness and are vulnerable to the curse of dimensionality.

Our second contribution is a unified statistical inference procedure based on the weighted bootstrap, which avoids estimating the complex components in the efficient influence function as required in existing DML methods. 
We provide theoretical justification for the validity of the proposed inference method. Simulation results reveal that the proposed BNNW approach consistently outperforms existing
alternatives, particularly in small-sample settings.

This paper is organized as follows. Section~\ref{sec:Basic-Framework-and}
sets up the basic framework and Section~\ref{sec:Estimation Method} proposes the BNNW estimator. Section~\ref{sec:Theoretical-Properties} establishes $\sqrt{N}$-consistent, asymptotic normality and semiparametric efficiency
results for the BNNW estimator. In Section~\ref{sec:Inference},
we provide a statistical inference procedure by weighted bootstrap.
In Section \ref{sec:Monte-Carlo-studies}, we illustrate the advantages
of the BNNW estimator in a finite sample by Monte Carlo studies.
In Section \ref{sec:Real-data}, we apply the BNNW method
to two specific datasets---the 401(k) data and the Mother's Significant Features dataset---to assess its effectiveness in estimating treatment
effects. All proofs
are relegated to the Supplementary Materials (\citealp{Wu2025BNNWsupp}).

\textbf{Notation.} We maintain the following notation conventions
throughout the paper. The set of non-negative integers is denoted
by $\N:=\{0,1,2,\ldots\}$ and the set of positive integers is denoted
by $\N^{+}:=\{1,2,\ldots\}$. For any column vector $\bs x=\left(x_{1},x_{2},\ldots,x_{d}\right)^{\top}\in\mathbb{R}^{d}$,
where $\mathbb{R}^{d}$ is the $d$-dimensional Euclidean space, $\left\Vert \bs x\right\Vert =\left(\bs x^{\top}\bs x\right)^{1/2}$
denotes its Euclidean norm. For any matrix $A=\left(a_{ij}\right)_{n\times n}$,
$\left\Vert A\right\Vert $ denotes its maximum singular value, i.e.,
the operator norm, and $\lambda_{\text{min}}(A)$ denotes its minimum
singular value. Let $\left(\Omega,\mathcal{F},P\right)$ be a probability
space. For any random vector (or matrix) $\bs X\in\mathcal{X}\subset\mathbb{R}^{d}$,
define $\left\Vert \bs X\right\Vert _{P,q}:=\left(\int\left\Vert \bs X\right\Vert ^{q}dP\right)^{1/q}$.
For any function $f:\mathcal{X}\subset\R^{d}\to\R^{d^{\prime}}$,
define $\left\Vert f\right\Vert _{P,q}:=\left\Vert f\left(\bs X\right)\right\Vert _{P,q}$
and $\left\Vert f\right\Vert _{\infty}:=\sup_{\bs x\in\mathcal{X}}\left\Vert f(\bs x)\right\Vert $
denotes it infinity norm. For any real number $a$, we let $\left\lceil a\right\rceil :=\min\left\{ b\in\mathbb{Z}:b\geq a\right\} $.
Let $\tod$ denote convergence in distribution. For any set $D$,
$\left|D\right|$ denotes its cardinality. For two positive non-random
sequences $a_{n},b_{n}$ and a random vector sequence $X_{n}$, $X_{n}=o_{P}(a_{n})$
means $P\left(\left\Vert X_{n}\right\Vert /a_{n}>\epsilon\right)\to0$
as $n\to\infty$ for any $\epsilon>0$, and $X_{n}=O_{P}(a_{n})$
and $X_{n}\lesssim_{P}a_{n}$ both represent that for any $\epsilon>0$, there
exists $M>0$ such that $\limsup_{n\to\infty}P\left(\left\Vert X_{n}\right\Vert /a_{n}\geq M\right)<\epsilon$.
We write $a_{n}\lesssim b_{n}$ if $a_{n}\leq Cb_{n}$ for a positive
constant $C$ independent of $n$, and $a_{n}\asymp b_{n}$ if $a_{n}\lesssim b_{n}$
and $b_{n}\lesssim a_{n}$. For a set of functions $\mathcal{F}$, we define the
covering number $N\left(\epsilon,\mathcal{F},\left\Vert \cdot\right\Vert _{Q,q}\right)$
as the minimal number $N$ of functions $f_{1},\ldots,f_{N}$ such
that $\sup_{f\in\mathcal{F}}\inf_{i=1,\ldots,N}\left\Vert f_{i}-f\right\Vert _{Q,q}<\epsilon$.

\section{Settings and Objective}\label{sec:Basic-Framework-and}
Let $T$ be the observed treatment variable with probability distribution function $F_{T}(\cdot)$ on the support $\mathcal{T}\subset\R$, where $\mathcal{T}$ can be a discrete set, a continuum, or a mixture of the discrete and continuum subsets. We denote the potential outcome when the treatment variable takes the value $t$ as
$Y^{*}(t)\in \mathbb{R}$ and the observed outcome as $Y=Y^*(T)$.  Let $L(y,z):\mathbb{R}\times\R\to[0,\infty)$ be a general loss
function, whose partial derivative with respect to $z$, denoted
by $L^{\prime}(y,z):=\partial L(y,z)/\partial z$, exists almost everywhere.  The treatment effect parameter of interest $\bs{\beta}^{*}=(\beta_{0}^{*},\beta_{1}^{*},\ldots,\beta_{p-1}^{*})^{\top}$
is defined by the unique solution of the following optimization problem:
\begin{equation}
\bs{\beta}^{*}:=\underset{\bs{\beta}\in\R^{p}}{\arg\min}\int_{\mathcal{T}}\e\left[L(Y^{*}(t),g(t;\bs{\beta}))\right]dF_{T}(t),\label{eq:def of the casual efffect}
\end{equation}
where $g(\cdot;\bs{\beta})$ is a parametric dose-response function indexed
by a $p$-dimensional parameter $\bs{\beta}=(\beta_{0},\beta_{1},...,\beta_{p-1})^{\top}\in\R^{p}$
for some $p\in\N^{+}$. We assume that $g(t;\boldsymbol{\beta})$
is twice differentiable with respect to $\boldsymbol{\beta}$. In particular, for discrete treatment $\mathcal{T}=\{0,1,\ldots,J\}$ for some integer $J\geq 1$, $g(t;\bs{\beta})=\sum^J_{j=0}\beta_j 1(t=j)$. The formulation (\ref{eq:def of the casual efffect}) is called the \emph{general treatment model}, since it covers a broad class of treatment effect models (see Table~\ref{tab:TreatmentModels}). 

\begin{table}[t]
    \centering
\begin{tabular}{@{}ll@{}}
\toprule
\multicolumn{1}{c}{$L(y,z)$}                        & \multicolumn{1}{c}{Resulted Treatment Effect Parameter}   \\ \midrule
$(y-z)^2/2$                                         & $g(t;\bs{\beta}^*)=\mathbb{E}\{Y^*(t)\}$                  \\
$(y-z)\cdot\{\tau-1(y-z\leq0)\}$ for $\tau\in(0,1)$ & $g(t;\bs{\beta}^*)$ is the $\tau$-th quantile of $Y^*(t)$ \\
$-y\log z-(1-y)\log(1-z)$ for $y\in\{0,1\}$         & $g(t;\bs{\beta}^*) = P\{Y^*(t)=1\}$                       \\ \bottomrule
\end{tabular}
    \caption{Example treatment effect models covered by \eqref{eq:def of the casual efffect}, which includes average dose-response function (row 1) \citep[e.g.,][]{cattaneo2010efficient, kennedy2017Nonparametric}, quantile dose-response function (row 2), and the success rate of potential outcomes (row 3). For binary $\mathcal{T}=\{0,1\}$, taking $g(t,\bs{\beta}) = (1-t)\beta_0 + t\beta_1$, we have $\beta^*_1-\beta^*_0$ for the cases in rows~1 and 2 correspond to the average treatment effect (ATE) studied in \cite{hahn1998Role, hirano2003Efficient} and the quantile treatment effect (QTE) in \cite{firpo2007Efficient}, respectively.}
    \label{tab:TreatmentModels}
\end{table}

To identify the treatment effect parameter associated with the unobservable potential outcome $Y^{*}(t)$ in observational studies, one of the most widely used assumptions in causal inference is the unconfoundedness condition \citep{rosenbaum1983central}: 
\begin{assumption}
\label{assu:unconfounded}A set of covariates, $\bs X \in \mathbb{R}^d$ for some positive integer $d$, is observed such that, conditional on $\bs X$, the treatment $T$ is
independent of the potential outcome $Y^{*}(t)$ for all $t\in\mathcal{T}$;
that is $Y^{*}(t)\perp T\mid\bs X$ for all $t\in\mathcal{T}$.
\end{assumption}

This paper considers high-dimensional covariate $\bs{X}$, under the premise that Assumption~\ref{assu:unconfounded} is generally more plausible when the covariate dimension $d$ is large (see, e.g., \citealp{farrell2015Robust,fan2022Estimation,athey2018Approximate,su2019Nonseparable}). Under Assumption~\ref{assu:unconfounded}, we can show that $\bs{\beta}^{*}$ is identified as the solution of the following
weighted optimization problem: 
\begin{equation}
\bs{\beta}^{*}=\underset{\bs{\beta}\in\R^{p}}{\arg\min}\ \e\left[\pi_{0}(T,\bs X)L(Y,g(T;\bs{\beta}))\right],\label{eq:indentification of beta0}
\end{equation}
where $\pi_{0}(T,\bs X)$ is the \emph{stabilized weighting function}
defined by 
\begin{equation}
\pi_{0}(T,\bs X):=\frac{dF_{T}(T)}{dF_{T|X}(T\mid\bs X)},\label{eq:desity ratio expression}
\end{equation}
and $F_{T|X}(\cdot)$ denotes the conditional distribution function
of $T$ given $\bs X$. 

\subsection{Problem \label{sec:problem}}
Estimating nuisance parameters, such as the stabilized weighting function $\pi_0$, is essential for the treatment effect analysis. While parametric methods are prone to model misspecification, nonparametric methods introduce bias into the final estimator of the treatment effect parameter $\bs{\beta}^*$ and suffer from the curse of dimensionality. Our goal is to develop a robust, efficient, and computationally feasible method for estimating and inferring $\bs{\beta}^*$ for high-dimension $d$ using a random sample $\left\{( Y_{i},\bs X_{i},T_{i})\right\} _{i=1}^n$ drawn from
the joint distribution of $\left(Y,\bs X,T\right)$. To motivate our approach, we first review existing results and highlight the problems we want to solve.

Let $h(Y,T;\bs{\beta}):=L^{\prime}(Y,g(T;\boldsymbol{\beta}))\cdot\partial g(T;\bs{\beta})/\partial\bs{\beta}$, so that $\pi_0(T,\bs{X})h(Y,T;\bs{\beta})$ is the score function for $\bs{\beta}^*$, and $\mathcal{L}(\cdot)$ be some convex
function with a well-defined derivative $\mathcal{L}^{\prime}(\cdot)$. Consider the special case where $L(y,z)=\mathcal{L}(y-z)$. Under Assumption \ref{assu:unconfounded},
\citet[Theorem 1]{ai2021Unified} derived the semiparametric efficient influence function (EIF)
for $\bs{\beta}^{*}$  \citep[see][for detailed introduction of EIF]{bickel1993Efficient}. Specifically, any efficient estimator $\widehat{\bs{\beta}}_{\text{eff}}$ of $\bs{\beta}^{*}$ satisfies
\begin{equation*}
\widehat{\bs{\beta}}_{\text{eff}} - \bs{\beta}^* =-\frac{1}{n}\sum^n_{i=1} \Sigma_{0}^{-1}\psi(Y,T,\boldsymbol{X};\boldsymbol{\beta}^{*}, \pi_0, \mu_0(T,\bs{X};\bs{\beta}^*)) + o_P(n^{-1/2})\,,
\end{equation*}
where $\Sigma_{0}:=\partial\mathbb{E}\left[\pi_{0}(T,\boldsymbol{X})h(Y,T;\bs{\beta})\right]\big/\partial\bs{\beta}^\top\big|_{\boldsymbol{\beta}=\boldsymbol{\beta}^{*}}$, and
\begin{align}
 & \psi(Y,T,\boldsymbol{X};\boldsymbol{\beta}^{*},\pi_{0},\mu_{0}(T,\bs{X};\bs{\beta}^{*})):=\pi_{0}(T,\boldsymbol{X})h(Y,T;\bs{\beta}^{*})-\pi_{0}(T,\boldsymbol{X})\mu_{0}(T,\boldsymbol{X};\boldsymbol{\beta}^{*})\notag\\
 & \qquad\qquad\qquad+\mathbb{E}\left[\mu_{0}(T,\boldsymbol{X};\boldsymbol{\beta}^{*})\pi_{0}(T,\boldsymbol{X})\mid T\right]+\mathbb{E}\left[\mu_{0}(T,\boldsymbol{X};\boldsymbol{\beta}^{*})\pi_{0}(T,\boldsymbol{X})\mid\boldsymbol{X}\right]\label{def:EIF}
\end{align}
with $\mu_{0}(T,\boldsymbol{X};\boldsymbol{\beta}):=\mathbb{E}[h(Y,T;\bs{\beta})\mid T,\boldsymbol{X}]$.
\cite{ai2021Unified} further proposed an efficient estimator for $\bs{\beta}^*$ for the case where $L(y,z)=\mathcal{L}(y-z)$. However, their estimator relies on a nonparametric linear sieve approximation of $\pi_0$, introducing bias and making their method behave poorly for even moderate dimensional covariate $\bs{X}$. 

To overcome these problems, double/debiased machine learning
(DML) techniques that effectively reduce bias have been explored for binary ATE and QTE models \citep{chernozhukov2018Double,kallus2022Localized}, which are special cases of our general treatment model \eqref{eq:def of the casual efffect} (see Table~\ref{tab:TreatmentModels}). To review their approach, we consider the case where $L(y,z) = \mathcal{L}(y-z)$, and illustrate how their methodology can be extended to our broader class of treatment models.
In this context, the DML estimator $\widehat{\bs{\beta}}_{\text{DML}}$ is defined as the solution to the
efficient influence function/score equation:
\begin{equation}
0 = \sum^n_{i=1}\psi(Y_i, T_i, \boldsymbol{X}_i;\bs{\beta}, \pi_0, \mu_0(T_i,\bs{X}_i;\bs{\beta}))\label{eq:estimatingEQ}
\end{equation}
for $\bs{\beta}$, where $\psi$ is defined in \eqref{def:EIF}.
The key driver of DML lies in the orthogonality of the EIF
with respect to the nuisance functions. Specifically, the derivatives of $\e[\psi(Y_i, T_i, \boldsymbol{X}_i;\bs{\beta}, \pi_0, \mu_0(T_i,\bs{X}_i;\bs{\beta}))]$ with respect to $\pi_0$ and $\mu_0$ vanishes at $\bs{\beta} = \bs{\beta}^{*}$. This orthogonality makes the estimating equation \eqref{eq:estimatingEQ} robust to small errors in estimating $\pi_0$ and $\mu_0$, as their impact enters only through higher-order terms, thereby reducing the bias in estimating $\bs{\beta}^*$. 
%
Following the localization idea in \cite{kallus2022Localized}, it can be shown that solving \eqref{eq:estimatingEQ} with $\bs{\beta}$ in $\mu_0$ replaced with $\bs{\beta}^*$  also yields a debiased estimator for $\bs{\beta}^*$, and the localized DML estimator in our context is $\widehat{\bs{\beta}}_{\text{LDML}}$ that solves
\begin{equation}
\begin{aligned}0= & \sum_{i=1}^{n}\widehat{\pi}(T_{i},\boldsymbol{X}_{i})h(Y_{i},T_{i};\widehat{\bs{\beta}}_{\text{LDML}})-\sum_{i=1}^{n}\widehat{\pi}(T_{i},\boldsymbol{X}_{i})\widehat{\mu}(T_{i},\boldsymbol{X}_{i};\widehat{\bs{\beta}}_{\text{init}})\\
 & +\sum_{i=1}^{n}\mathbb{E}\left[\widehat{\mu}(T_{i},\boldsymbol{X}_{i};\widehat{\bs{\beta}}_{\text{init}})\widehat{\pi}(T_{i},\boldsymbol{X}_{i})\mid T_{i}\right]+\sum_{i=1}^{n}\mathbb{E}\left[\widehat{\mu}(T_{i},\boldsymbol{X}_{i};\widehat{\bs{\beta}}_{\text{init}})\widehat{\pi}(T_{i},\boldsymbol{X}_{i})\mid\boldsymbol{X}_{i}\right]\,,
\end{aligned}
\label{eq:dml for general treatment model}
\end{equation}
where $\widehat{\bs{\beta}}_{\text{init}}$ is an initial estimator of $\bs{\beta}^*$. 

The DML-based idea is hindered by at least two significant obstacles. 
The first one is that these methods
all estimate $\pi_0$ by the inverse of some estimator of the propensity score. However, estimating the propensity score is known to be difficult in the presence of high-dimensional confounders $\bs X$
and the inverse is sensitive to the estimated propensity score \citep{kang2007demystifying}, which makes results unstable, especially when the sample size is small. The second obstacle is
that when solving the equality (\ref{eq:dml for general treatment model}),
two additional conditional expectations, namely 
$\mathbb{E}\left[\widehat{\mu}(T,\boldsymbol{X};\widehat{\bs{\beta}}_{\text{init}})\widehat{\pi}(T,\bs X)\mid T\right]$
and $\mathbb{E}\left[\widehat{\mu}(T,\boldsymbol{X};\widehat{\bs{\beta}}_{\text{init}})\widehat{\pi}(T,\bs X)\mid\boldsymbol{X}\right]$,
must be estimated, which is highly undesirable from both a theoretical
and computational perspective. 

In the following section, we introduce our balanced neural networks weighting (BNNW) estimator for $\bs{\beta}^{*}$. In this procedure, $\pi_0$ is estimated directly, without taking the inverse of an estimated propensity score. Unlike existing DML approaches, the proposed estimator does not rely on the EIF, thereby avoiding the need to estimate the last two terms in \eqref{eq:dml for general treatment model}. Instead, it achieves debiased and robust estimation through a novel covariate balancing calibration. Moreover, by leveraging deep neural networks (DNNs), the estimator is well-suited for handling high-dimensional covariates, mitigating the ``curse of dimensionality''.

\section{Estimation Method \label{sec:Estimation Method}}

\begin{figure}[!htbp]
    \centering
        \includegraphics[width=1\textwidth]{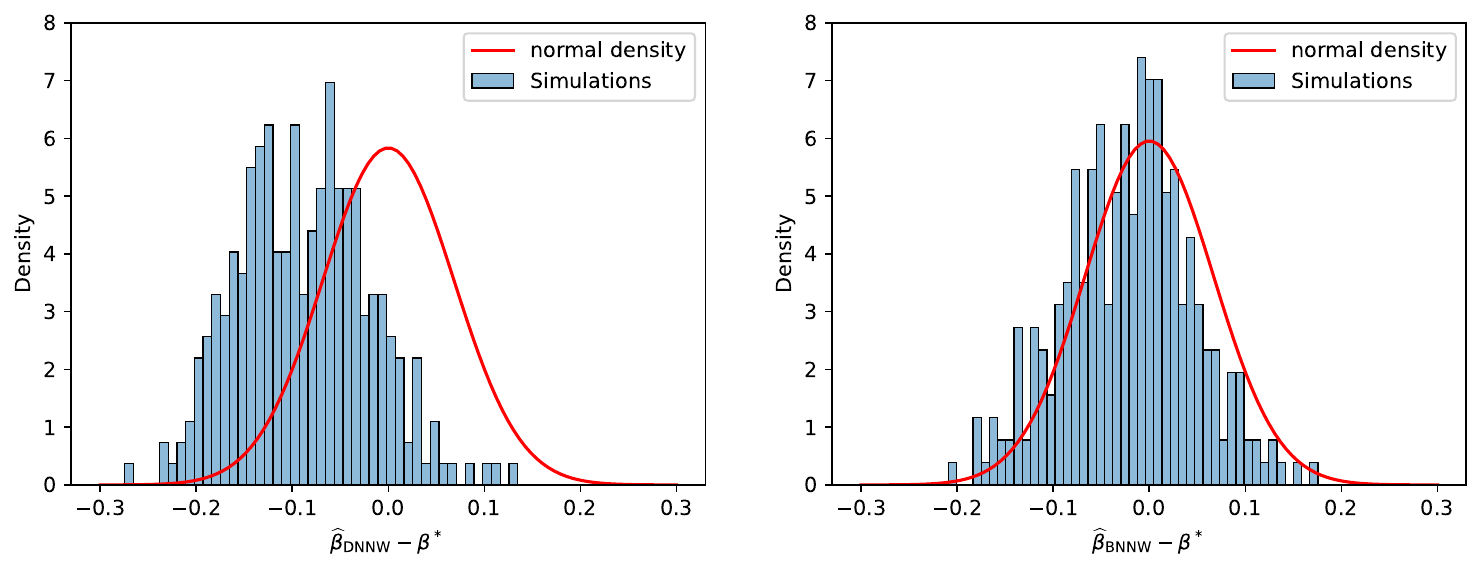}
	\caption{Comparison of the DNNW (left) and BNNW (right) estimators. Empirical distributions of the error of the estimators of the first component in the parameter $\bs{\beta}$, from $300$ Monte Carlo replications of samples under a continuous treatment setting with a sample size $N=2000$ and $d=20$ covariates. The dose-response model $g(t;\bs{\beta})$ is a linear polynomial of a continuous treatment $t$.}
	\label{fig:compare_dnn_bnn}
\end{figure}

This section proposes our novel BNNW method, which consists of two steps. In Section~\ref{sec:The-estimation-of}, we first construct an initial deep neural networks weighting (DNNW) estimator $\widehat{\bs{\beta}}_{\mathrm{DNNW}}$ by minimizing the weighted sample average of $L(Y_i,g(T_i;\bs{\beta}))$, where the stabilized weighting function $\pi_0(T_i,\bs{X}_i)$ is estimated by a DNN, denoted $\widehat{\pi}_{\mathrm{DNN}}(T_i,\bs{X}_i)$. As shown in the left panel of Figure~\ref{fig:compare_dnn_bnn}, like many methods with machine learning-based nuisance estimators discussed in \citet{chernozhukov2018Double}, the DNNW estimator $\widehat{\bs{\beta}}_{\mathrm{DNNW}}$ inherits the substantial bias of $\widehat{\pi}_{\mathrm{DNN}}$. To mitigate this bias, Section \ref{sec:bell} introduces a BNN estimator, $\widehat{\pi}_{\mathrm{BNN}}$,  which reweights $\widehat{\pi}_{\mathrm{DNN}}$ using the covariate balancing property of $\pi_0$ on a specific instrument function. The resulting estimator $\widehat{\bs{\beta}}_{\mathrm{BNNW}}$, obtained by minimizing the weighted sample average of $L(Y_i, g(T_i;\bs{\beta}))$ with weights $\widehat{\pi}_{\mathrm{BNN}}(T_i,\bs{X}_i)$, substantially reduces bias; see the right panel of Figure~\ref{fig:compare_dnn_bnn} for illustration and a formal proof is deferred to Section~\ref{sec:Theoretical-Properties}. To avoid potential over-fitting due to dependence between the initial and final estimators, Section \ref{sec:cross-fitting} details the implementation of BNNW using a cross-fitting strategy.

\subsection{Deep Neural Networks Initial Estimation\label{sec:The-estimation-of}}

Based on \eqref{eq:indentification of beta0},
a natural method of estimating $\bs{\beta}^{*}$ is first estimating
the nuisance function $\pi_{0}(T,\bs X)$ nonparametrically, denoted
by $\widehat{\pi}(T,\bs X)$, then constructing the estimator of $\bs{\beta}^{*}$
by 
\begin{equation}
\widehat{\bs{\beta}}=\underset{\bs{\beta}\in\R^{p}}{\arg\min}\sum_{i=1}^n\widehat{\pi}(T_{i},\bs X_{i})L(Y_{i},g(T_{i};\bs{\beta})).\label{eq:naive minimization step}
\end{equation}
%
%
To estimate $\pi_{0}$ as a whole with a DNN, we note that $\pi_{0}$ can be identified as the unique solution of the following minimization problem:
\begin{align}
\pi_{0}(\cdot,\cdot) & =\underset{\pi}{\arg\min}\,\e\left[(\pi(T,\bs X)-\pi_{0}(T,\bs X))^{2}\right]\nonumber \\
& =\underset{\pi}{\arg\min}\left\{ \e\left[\pi^{2}(T,\bs X)\right]-2\int_{\mathcal{T}}\int_{\mathcal{X}}\pi(t,\bs x)dF_{T}(t)dF_{X}(\bs x)+\e\left[\pi_{0}^{2}(T,\bs X)\right]\right\} \nonumber \\
&=\underset{\pi}{\arg\min}\left[\e\left\{\pi^{2}(T,\bs X)\right\}-2\int_{\mathcal{T}}\int_{\mathcal{X}}\pi(t,\bs x)dF_{T}(t)dF_{X}(\bs x)\right] =:\underset{\pi}{\arg\min}\,R(\pi),\label{eq:population minimize of pi}
\end{align}
where the definition of $R(\pi)$ is clear.

Let  $\f_{\mathrm{DNN}}(\mathcal{W},\mathcal{D})$ be a class of transformed DNNs with depth $\mathcal{D}$
and width $\mathcal{W}$ defined as:
\begin{align*}
& {\f_{\mathrm{DNN}}(\mathcal{W},\mathcal{D})=  \Bigl\{\bs z\in\R^{d+1}\mapsto\phi\left( W_{\mathcal{D}}\boldsymbol{\sigma}\Big(\cdots\boldsymbol{\sigma}\big(W_{1}\boldsymbol{\sigma}(W_{0}\boldsymbol{z}+b_{0})+b_{1}\big)+\cdots\Big)+b_{\mathcal{D}}\right) :}\\
&{\qquad\quad W_{l}\in\R^{d_{l+1}\times d_{l}},b_{l}\in\R^{d_{l+1}},0\leq l\leq\mathcal{D},d_{0}=d+1,d_{\mathcal{D}+1}=1,\max\{d_{1},\ldots,d_{\mathcal{D}}\}\leq\mathcal{W}\Bigl\}\,,}
\end{align*}
where $\sigma(z):=\max\{z,0\}$ is the rectified linear unit (ReLU) activation function for $z\in\mathbb{R}$, $\boldsymbol{\sigma}:\mathbb{R}^{d_l}\to \mathbb{R}^{d_l}$ denotes the function applying $\sigma(\cdot)$ componentwisely, and $\phi(\cdot):\mathbb{R}\to [0,\infty)$ is a user-specified transformation function such that  the output of the neural network  $\f_{\mathrm{DNN}}(\mathcal{W},\mathcal{D})$ is non-negative. In simulation studies and real applications, we  adopt the exponential function $\phi(z)=\exp(z)$.  We refer to \citet[Section 2.1]{farrell2021Deepa} for more discussion on the construction of a DNN.

Based on \eqref{eq:population minimize of pi}, the initial DNN estimator of $\pi_{0}$
is defined by 
\begin{equation}
\widehat{\pi}_{\mathrm{DNN}}=\underset{\pi\in\f_{\mathrm{DNN}}(\mathcal{W},\mathcal{D})
}{\arg\min}\widehat{R}(\pi),\label{eq:estimation of pi definition}
\end{equation}
where 
\[
\widehat{R}(\pi):=\frac{1}{n}\sum_{i=1}^n\pi^{2}(T_{i},\bs X_{i})-\frac{2}{n(n-1)}\sum_{j=1}^{n}\sum_{l=1,l\neq j}^{n}\pi(T_{j},\bs X_{l})\,.
\]
%
Then the DNNW estimator of $\bs{\beta}^{*}$ is defined by 
\begin{align}
\widehat{\bs{\beta}}_{\mathrm{DNNW}}=\underset{\bs{\beta}\in\R^{p}}{\arg\min}\sum_{i=1}^n\widehat{\pi}_{\mathrm{DNN}}(T_{i},\bs X_{i})L(Y_{i},g(T_{i};\bs{\beta})).\label{def:beta_DNN}
\end{align}

\subsection{Debias using Balanced Neural Networks Weighting\label{sec:bell}}

This section introduces the main idea of our debiased BNNW method. First, note that $\pi_0$ satisfies the following covariate balancing property: 
\begin{equation}
\e\left[\pi_{0}(T,\bs X)\bs{\xi}(T,\bs X)\right]=\int_{\mathcal{X}}\int_{\mathcal{T}}\bs{\xi}(t,\bs x)dF_{T}(t)dF_{X}(\bs x)\label{eq:property of pi0}
\end{equation}
for any integrable function $\bs{\xi}:\mathcal{T}\times\mathcal{X}\mapsto\mathbb{R}^{d_\xi}$ for some integer $d_\xi\geq 1$.\footnote{It shows that $\pi_0$ balances the confounders across different treatment levels. Take the binary treatment case as an example, \eqref{eq:property of pi0} is equivalent to $\e\{\pi_0(1,\bs{X})u(\bs{X})T\} = \e\{\pi_0(0,\bs{X})u(\bs{X})(1-T)\}=\e\{u(\bs{X})\}$ for any integrable function $u(\cdot)$.}
A good estimator of $\pi_{0}(T,\bs X)$ (debiases for estimating $\bs{\beta}^*$ or not) should satisfy the empirical analogue of \eqref{eq:property of pi0}.
Second, we found that selecting an appropriate $\bs{\xi}(T,\bs X)$ can give a debiased estimator for $\bs{\beta}^{*}$. However, the construction of the DNN estimator $\widehat{\pi}_{\mathrm{DNN}}$ defined in \eqref{eq:estimation of pi definition} does not guarantee these. 
We propose to calibrate the DNN estimator $\widehat{\pi}_{\mathrm{DNN}}$ based on \eqref{eq:property of pi0} as follows. 

Let $D(\cdot,\cdot):\mathbb{R}^2\to \R$ be a discrepancy measure such that $D(v):=D(v,1)$ is twice continuously
differentiable and strictly convex in $v\in\R$, i.e. $ D'(1)=0$
and $D''(1)>0$. With a given instrumental
function $\bs{\xi}(t,\bs x)$ and the initial estimator
$\widehat{\pi}_{\mathrm{DNN}}$, we define the BNN estimator of $\pi_0$ as $$\widehat{\pi}_{\mathrm{BNN}}(T_{i},\bs X_{i}):=\widehat{w}_{i}\cdot\widehat{\pi}_{\mathrm{DNN}}(T_{i},\bs X_{i}) \ \text{for}\ i\in \{1,...,n\}\,,$$ where the calibration weights $(\widehat{w}_1,\ldots,\widehat{w}_n)$ solve the following optimization problem:
\begin{align}
\begin{cases}
\min_{w_{i}:1\leq i\leq n}\sum_{i=1}^nD(w_{i}) \ \text{subject to}\\
\frac{1}{n}\sum_{i=1}w_{i}\widehat{\pi}_{\mathrm{DNN}}(T_{i},\bs X_{i})\bs{\xi}(T_{i},\bs X_{i})=\frac{1}{n(n-1)}\sum_{j=1}^{n}\sum_{l=1,l\neq j}^{n}\bs{\xi}(T_{j},\bs X_{l}).
\end{cases}\label{eq:intro empirical constraints}
\end{align}

In Section \ref{sec: proof of EL calibration step} of the Appendix, we show that the dual solution of~\eqref{eq:intro empirical constraints} is given
by\footnote{\textcolor{black}{In our numerical studies, we take $D(v)=v\log v -v$, i.e., $\rho^{\prime}(v)=e^{-v}$, ensuring the weights $\widehat{w}_i$, $i=1,\ldots,n$, always non-negative, and preventing pathological solutions of \eqref{def:betabell} due to negative weighting.}}
\[
\widehat{w}_{i}=\rho^{\prime}\left\{\widehat{\pi}_{\mathrm{DNN}}(T_{i},\bs X_{i})\hat{\bs{\lambda}}^{\trans}\bs{\xi}(T_{i},\bs X_{i})\right\}\ \text{for}\ i\in \{1,...,n\},
\]
where $\rho(v):=D\left\{ (D^{\prime})^{-1}(-v)\right\} +v\cdot(D^{\prime})^{-1}(-v)$ is a strictly concave function,  $\rho^{\prime}(v)$ is the derivative of $\rho(v)$, and $\hat{\bs{\lambda}}$ is the unique maximizer of  the following concave function 
\[
\hat{G}(\bs{\lambda}):=\frac{1}{n}\sum_{i=1}^n\rho\{\widehat{\pi}_{\mathrm{DNN}}(T_{i},\bs X_{i})\bs{\lambda}^{\trans}\bs{\xi}(T_{i},\bs X_{i})\}-\frac{1}{n(n-1)}\sum_{j=1}^{n}\sum_{l=1,l\neq j}^{n}\bs{\lambda}^{\trans}\bs{\xi}(T_{j},\bs X_{l})\,.
\]
Then, the BNNW estimator of $\bs{\beta}^{*}$ is 
\begin{align}
\widehat{\bs{\beta}}_{\mathrm{BNNW}}=\underset{\bs{\beta}\in\R^{p}}{\arg\min}\ \sum_{i=1}^n\widehat{\pi}_{\mathrm{BNN}}(T_{i},\bs X_{i})L(Y_{i},g(T_{i};\bs{\beta})).\label{def:betabell}
\end{align}

The remaining problem is how to choose the instrumental function $\bs{\xi}(t,\bs x)$ to ensure $\widehat{\bs{\beta}}_{\mathrm{BNNW}}$ a debiased estimator. Let $\mu_{0}(t,\bs x;\bs{\beta}^{*}):=\e\left[L^{\prime}(Y,g(T;\bs{\beta}^{*}))\cdot\partial g(T;\bs{\beta}^{*})/\partial\bs{\beta}|T=t,\bs X=\bs x\right]$.
 In Appendix~\ref{sec: intuition on the choice of xi}, we illustrate that, for any integrable  $\bs{\xi}$, the bias of the resulting $\widehat{\bs{\beta}}_{\mathrm{BNNW}}$ can be bounded by $\left\Vert \widehat{\pi}_{\mathrm{BNN}}(T_{i},\bs X_{i})-\pi_{0}(T_{i},\bs X_{i})\right\Vert _{P,2}\times\inf_{\mathbf{Q}\in\mathbb{R}^{p\times d_{\xi}}}\left\Vert \mu_{0}(T,\bs X;\bs{\beta}^{*})-\mathbf{Q}\bs{\xi}(T,\bs X)\right\Vert_{P,2}$, and $\left\Vert \widehat{\pi}_{\mathrm{BNN}}(T_{i},\bs X_{i})-\pi_{0}(T_{i},\bs X_{i})\right\Vert _{P,2}\lesssim \left\Vert \widehat{\pi}_{\mathrm{DNN}}(T_{i},\bs X_{i})-\pi_{0}(T_{i},\bs X_{i})\right\Vert _{P,2}$ holds under suitable conditions. Thus, in order to eliminate bias, the most desirable choice is $\bs{\xi}(t,\bs x)=\mu_{0}(t,\bs x;\bs{\beta}^{*})$,
which is unknown in practice.
~By Taylor's expansion, $\mu_{0}(t,\bs x;\bs{\beta}^{*})\approx\mu_{0}(t,\bs x;\widehat{\bs{\beta}}_{\mathrm{DNNW}})+\partial_{\bs{\beta}}\mu_{0}(t,\bs x;\widehat{\bs{\beta}}_{\mathrm{DNNW}})(\bs{\beta}^{*}-\widehat{\bs{\beta}}_{\mathrm{DNNW}})$. We thus consider the following instrumental function: 
\begin{align*}
\widehat{\bs{\xi}}(t,\bs x)=\begin{pmatrix}\widehat{\mu}(t,\bs x;\widehat{\bs{\beta}}_{\mathrm{DNNW}})\\
\vec{\left\{ \widehat{\partial_{\bs{\beta}}\mu}(t,\bs x;\widehat{\bs{\beta}}_{\mathrm{DNNW}})\right\} }
\end{pmatrix},
\end{align*}
where $\widehat{\mu}(t,\bs x;\widehat{\bs{\beta}}_{\mathrm{DNNW}})$ and $\widehat{\partial_{\bs{\beta}}\mu}(t,\bs x;\widehat{\bs{\beta}}_{\mathrm{DNNW}})$
are some machine learning-based estimators of $\mu_{0}(t,\bs x;\widehat{\bs{\beta}}_{\mathrm{DNNW}})$
and $\partial_{\bs{\beta}}\mu_{0}(t,\bs x;\widehat{\bs{\beta}}_{\mathrm{DNNW}})$
respectively, and $\vec{\left\{ \widehat{\partial_{\bs{\beta}}\mu}(t,\bs x;\widehat{\bs{\beta}}_{\mathrm{DNNW}})\right\} }$
denotes the vector collecting all elements of the matrix $\widehat{\partial_{\bs{\beta}}\mu}(t,\bs x;\widehat{\bs{\beta}}_{\mathrm{DNNW}})$.\footnote{Without loss of generality, we assume that the elements in $\widehat{\bs{\xi}}(t,\bs x)$ are linearly independent. Otherwise, we can remove the redundant elements from $\widehat{\bs{\xi}}(t,\bs x)$, which will not affect the results.} Specifically, $\widehat{\mu}(t,\bs x;\widehat{\bs{\beta}}_{\mathrm{DNNW}})$ and $\widehat{\partial_{\bs{\beta}}\mu}(t,\bs x;\widehat{\bs{\beta}}_{\mathrm{DNNW}})$ can be constructed by either state-of-the-art machine learning methods or regularized approaches, depending on the situation; See Section~\ref{sec:nuisance estimation} in the supplementary material for construction of estimates of $\mu_0$ and $\partial_{\bs{\beta}}\mu_0$ for different treatment models.

\subsection{Full BNNW Algorithm\label{sec:cross-fitting}}

Building on the procedures in Sections~\ref{sec:The-estimation-of} and~\ref{sec:bell}, we now present the complete BNNW algorithm. To incorporate flexible machine learning estimators $\widehat{\mu}_{0}(t,\bs x;\widehat{\bs{\beta}}_{\mathrm{DNNW}})$
and $\partial_{\bs{\beta}}\widehat{\mu}_{0}(t,\bs x;\widehat{\bs{\beta}}_{\mathrm{DNNW}})$, while avoiding potential overfitting, we employ cross-fitting. 

Specifically, we partition the sample into $K\geq 2$ equal folds. For simplicity, from now on, we denote the total sample size by $N$, and assume $N=nK$ for some $n\in\mathbb{N}^+$. Let $I_k = \{(k-1)\cdot n +1,\ldots, kn\}$ for $k=1,\ldots, K$ be 
the indices of the $k$-th fold, and $I_{-k}=\{1,\ldots,N\}\backslash I_{k}$
be the indices of the remaining data excluding the $k$-th fold.
The proposed BNNW algorithm proceeds as follows.
\begin{myalgorithm}[BNNW Algorithm]\label{alg:BNNW} 
\begin{enumerate}
\item for every $k=1,\ldots,K$, 
\begin{enumerate}
\item using the sample in $I_{-k}$ to construct an estimator $\widehat{\pi}_{\mathrm{DNN}}^{(-k)}(\cdot,\cdot)$
following \eqref{eq:estimation of pi definition};
\item splitting $I_{-k}$ into two parts $I_{-k}^{(1)}$ and $I_{-k}^{(2)}$
with equal size; using the sample in $I_{-k}^{(1)}$ to construct
an initial estimator $\widehat{\bs{\beta}}_{\mathrm{DNNW}}^{(-k)}$ following
\eqref{def:beta_DNN}; using the sample in $I_{-k}^{(2)}$ to construct
estimators $\widehat{\mu}^{(-k)}(\cdot,\cdot;\widehat{\bs{\beta}}_{\mathrm{DNNW}}^{(-k)})$
and $\widehat{\partial_{\bs{\beta}}\mu}^{(-k)}(\cdot,\cdot;\widehat{\bs{\beta}}_{\mathrm{DNNW}}^{(-k)})$
respectively;
\item Calculate the solution of the proposed balanced calibration problem: 
\[
\widehat{w}^{(k)}_{i}=\rho^{\prime}\left(\widehat{\pi}^{(-k)}_{\mathrm{DNN}}(T_{i},\bs X_{i})\hat{\bs{\lambda}}_k^{\trans}\widehat{\bs{\xi}}^{(k)}(T_{i},\bs X_{i})\right)\ \text{for}\ i\in I_k,
\]
where $\hat{\bs{\lambda}}_k$ is the unique maximizer of the refined concave function 
\[
\hat{G}(\bs{\lambda}):=\frac{1}{n}\sum_{i\in I_k}\rho\left\{\widehat{\pi}^{(-k)}_{\mathrm{DNN}}(T_{i},\bs X_{i})\bs{\lambda}^{\trans}\widehat{\bs{\xi}}^{(k)}(T_{i},\bs X_{i})\right\}-\frac{1}{n(n-1)}\sum_{j\neq l \in I_{k}}\bs{\lambda}^{\trans}\widehat{\bs{\xi}}^{(k)}(T_{j},\bs X_{l})\,,
\]
with $\hat{\bs{\xi}}^{(k)}(t,\bs x):=\left(\widehat{\mu}^{(-k)}(t,\bs x;\widehat{\bs{\beta}}_{\mathrm{DNNW}}^{(-k)})^{\trans},\vec{\left\{ \widehat{\partial_{\bs{\beta}}\mu}^{(-k)}(t,\bs x;\widehat{\bs{\beta}}_{\mathrm{DNNW}}^{(-k)})\right\} }^{\trans}\right)^{\trans}$.
Then obtaining the BNN weights $\widehat{\pi}_{\mathrm{BNN}}^{(k)}(T_{i},\bs X_{i})=\widehat{w}_{i}^{(k)}\cdot\widehat{\pi}_{\mathrm{DNN}}^{(-k)}(T_{i},\bs X_{i})$
for $i\in I_{k}$; 
\item solving the optimization problem: 
\begin{equation}
\widehat{\bs{\beta}}_{\mathrm{BNNW}}^{(k)}=\underset{\bs{\beta}\in\R^{p}}{\arg\min}\sum_{i\in I_{k}}\widehat{\pi}_{\mathrm{BNN}}^{(k)}(T_{i},\bs X_{i})L(Y_{i},g(T_{i};\bs{\beta})).\label{eq:minimization step}
\end{equation}
\end{enumerate}
\item estimating $\bs{\beta}^{*}$ by 
\begin{equation}
\widehat{\bs{\beta}}_{\mathrm{BNNW}}=\frac{1}{K}\sum_{k=1}^{K}\widehat{\bs{\beta}}_{\mathrm{BNNW}}^{(k)}.\label{eq:main estimator}
\end{equation}
\end{enumerate}
\end{myalgorithm}

\begin{figure}[!h]
\centering \includegraphics[width=0.90\textwidth]{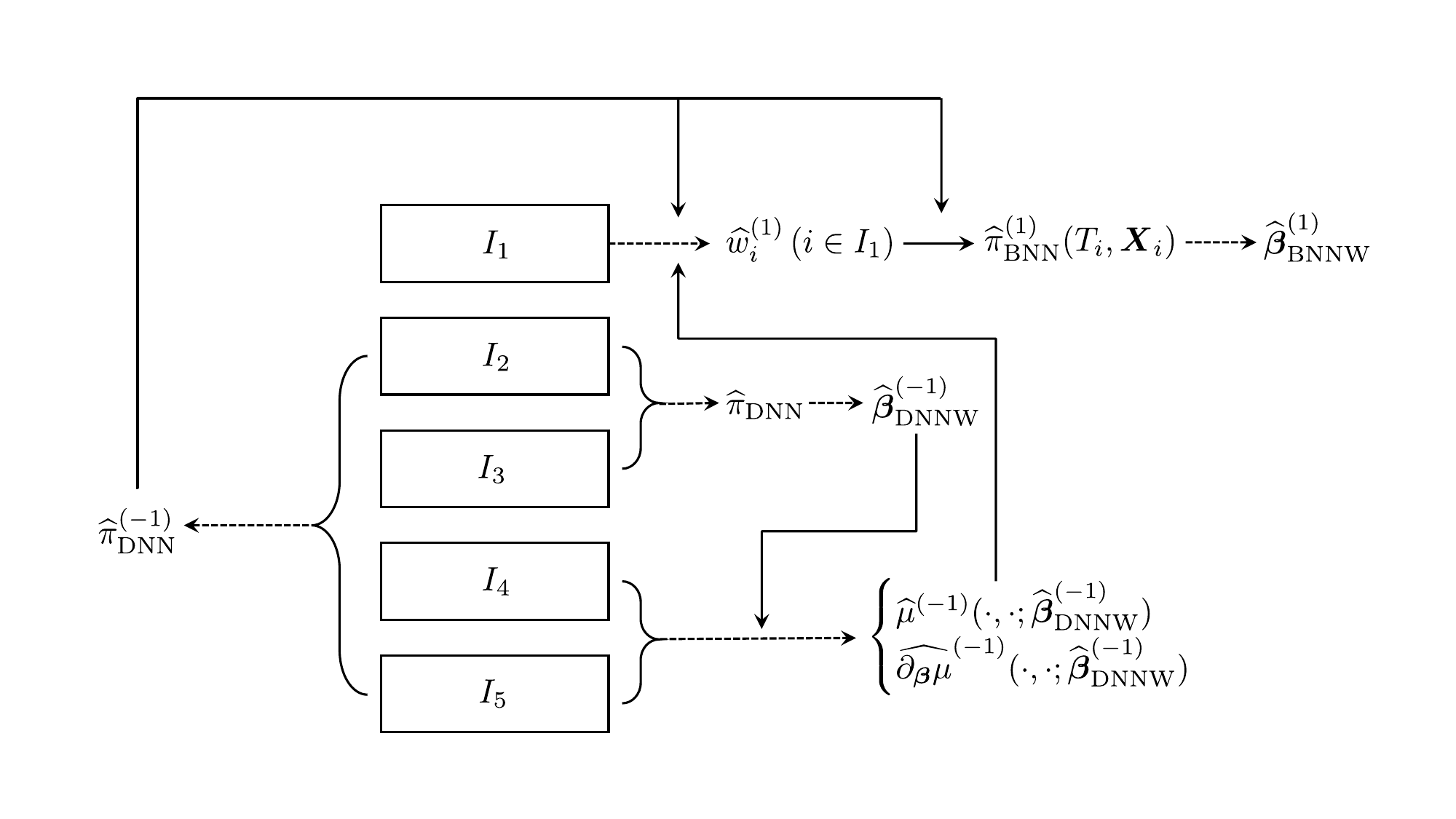}
\caption{Sketch of our BNNW algorithm when $K=5$ and $k=1$. Arrows
with dashed line denote estimation and arrows with solid line denote
plugging in.}
\label{fig:cross-fitting}
\end{figure}

\begin{rem}
The BNNW method takes a similar sample-splitting and cross-fitting
approaches as \citet{kallus2022Localized}. Their empirical results suggest that setting $K=5$ provides sufficiently good performance. Figure \ref{fig:cross-fitting} sketches the BNNW algorithm
for $\widehat{\bs{\beta}}_{\mathrm{BNNW}}^{(1)}$ when $K=5$.
\end{rem}

\section{Large Sample Properties\label{sec:Theoretical-Properties}}

\subsection{Convergence Rates of $\widehat{\pi}_{\mathrm{DNN}}$ and $\widehat{\bs{\beta}}_{\mathrm{DNNW}}$ \label{subsec:initial pi and betaDNN}}

This subsection establishes the convergence rates of the DNN
weights $\widehat{\pi}_{\mathrm{DNN}}$ and the initial estimator $\widehat{\bs{\beta}}_{\mathrm{DNNW}}$
defined in \eqref{eq:estimation of pi definition} and \eqref{def:beta_DNN}
respectively. Without loss of generality, we focus on the case that
$T$ is continuously distributed and assume that $\mathcal{X}=[0,1]^{d}$
and $\mathcal{T}=[0,1]$. The theory for discretely distributed $T$
can be similarly established.
\begin{assumption}
\label{assu:ass for estimate of pi} Let $M>1$ be a finite constant.
\begin{enumerate}[label=(\alph*)]
\item $\underset{(t,\bs x)\in[0,1]^{d+1}}{\inf}\pi_{0}(t,\bs x)\geq1/M$.
\item The stabilized weighting function $\pi_{0}(t,\bs x)$ lies in the
smooth function space $C^{s_{\pi}}\left([0,1]^{d+1}\right)$ with
smoothness $s_{\pi}\in\N^{+}$: 
\[
\pi_{0}(t,\bs x)\in C^{s_{\pi}}\left([0,1]^{d+1}\right):=\left\{ \pi:\max_{\bs{\alpha}\in\mathbb{\N}^{d+1},\left\Vert \bs{\alpha}\right\Vert _{1}\leq s_{\pi}}\underset{(t,\bs x)\in[0,1]^{d+1}}{\sup}\left|\partial^{\bs{\alpha}}\pi(t,\bs x)\right|\leq M\right\} ,
\]
where $\left\Vert \bs{\alpha}\right\Vert _{1}:=\alpha_{1}+\cdots+\alpha_{d+1}$
and $\partial^{\bs{\alpha}}\pi$ is the partial derivative.
\end{enumerate}
\end{assumption}
\begin{assumption}
\label{assu:regularity conditions} Recall $h(Y,T;\bs{\beta})=L^{\prime}(Y,g(T;\boldsymbol{\beta}))\cdot\partial g(T;\bs{\beta})/\partial\bs{\beta}$.
\begin{enumerate}[label=(\alph*)]
\item The parameter space $\varTheta\subset\R^{p}$ is a compact set and
the true parameter $\bs{\beta}^{*}$ is an interior point of $\varTheta$. 
\item For all $\delta>0$, there exists $\epsilon(\delta)>0$ such that
$\inf_{\left\Vert \bs{\beta}-\bs{\beta}^{*}\right\Vert >\delta}\left\Vert \e\left[\pi_{0}(T,\bs X)h(Y,T;\bs{\beta})\right]\right\Vert \geq\epsilon(\delta)>0$.
\item Define $\Sigma(\bs{\beta}):=\partial\e\left[\pi_{0}(T,\bs X)h(Y,T;\bs{\beta})\right]/\partial\bs{\beta}^{\trans}$
and $\Sigma_{0}:=\Sigma(\bs{\beta}^{*})$. The singular values of
$\Sigma_{0}$ are bounded between $c_{1}$ and $c_{2}$, where $0<c_{1}\leq c_{2}<\infty$
are constants. Furthermore, $\Sigma(\bs{\beta})$ is continuous at
$\bs{\beta}=\bs{\beta}^{*}$.
\end{enumerate}
\end{assumption}

\begin{assumption}
\label{assu:estimation of init DNN beta} Consider $\widehat{\bs\beta}_{\mathrm{DNN}}$ nearly minimizes the objective function in the sense that
\[
\left\Vert \frac{1}{n}\sum_{i=1}^{n}\widehat{\pi}_{\mathrm{DNN}}(T_{i},\bs X_{i})h(Y_{i},T_{i};\widehat{\bs{\beta}}_{\mathrm{DNNW}})\right\Vert \leq\left\Vert \frac{1}{n}\sum_{i=1}^{n}\widehat{\pi}_{\mathrm{DNN}}(T_{i},\bs X_{i})h(Y_{i},T_{i};\bs{\beta}^{*})\right\Vert +o_{P}(n^{-1/2}).
\]

\end{assumption}

\begin{assumption}
\label{assu:VC type class}For any $1\leq j\leq p$, let $h_{j}(Y,T;\bs{\beta})$
be the $j$-th component of $h(Y,T;\bs{\beta})$. The function class
$\mathcal{H}_{j}:=\left\{ h_{j}(Y,T;\bs{\beta}):\bs{\beta}\in\varTheta\right\} $
attached with the measurable envelop $H_{j}(Y,T):=\sup_{\bs{\beta}\in\varTheta}\left|h_{j}(Y,T;\bs{\beta})\right|$ has the covering number 
\[
\sup_{Q}\log N\left(\epsilon\left\Vert H_{j}\right\Vert _{Q,2},\mathcal{H}_{j},\left\Vert \cdot\right\Vert _{Q,2}\right)\leq\nu\log\left(a/\epsilon\right)\text{ uniformly over }0<\epsilon\leq1,
\]
where $a,\nu>0$ are constants and $\sup_{Q}$ is taken over all finitely
discrete probability measures. Furthermore, $\left\Vert \sup_{\bs{\beta}\in\varTheta}\left\Vert h(Y,T;\bs{\beta})\right\Vert \right\Vert _{P,2}<\infty$.
\end{assumption}
Assumption~\ref{assu:ass for estimate of pi} is a standard smoothness
condition for functional approximation, see also \citet{farrell2021Deepa}
and \citet{colangelo2023double}. The second part of this assumption
essentially implies the bounded restriction $\sup_{(t,\bs x)\in\mathcal{T}\times\mathcal{X}}\pi_{0}(t,\bs x)\leq M$. Assumptions~ \ref{assu:regularity conditions} and~\ref{assu:estimation of init DNN beta} are standard in M-estimation theory, see, e.g., \cite{chen2003Estimation,vaart1998Asymptotic}. Assumption~\ref{assu:VC type class} assumes $\mathcal{H}_{j}$ to
be VC type. Typically, this assumption is verified by showing that
$\mathcal{H}_{j}$ is a VC-subgraph class (\citet{vaart1996Weak}).
It can be shown that this assumption holds for a large variety of
the loss function $L(y,z)$ and the response model $g(t;\bs{\beta})$,
e.g., $g(t;\bs{\beta})=t\beta_{1}+(1-t)\beta_{0}$ with $L(y,z)=(y-z)^{2}/2$,
$L(y,z)=(y-z)(\tau-1(y-z\leq0))$ or $L(y,z)=(y-z)^{2}\left|\tau-1(y-z\leq0)\right|$.
Similar assumptions are also imposed by \citet{chernozhukov2018Double}
and \citet{kallus2022Localized}. 

\begin{proposition}
\label{thm:rate for init DNN beta}Suppose that Assumptions \ref{assu:unconfounded}-\ref{assu:VC type class}
hold. If the DNN class $\f_{\mathrm{DNN}}(\mathcal{W},\mathcal{D})$
is constructed with width $\mathcal{W}\geq17s_{\pi}^{d+2}3^{d+3}(d+1)$
and depth $\mathcal{D}\geq108s_{\pi}^{2}+2(d+1)$ satisfying $\mathcal{W}\mathcal{D}\asymp n^{(d+1)/(4s_{\pi}+2d+2)}\left(\log n\right)^{2}$ and $\left\Vert\widehat{\pi}_{\mathrm{DNN}}\right\Vert_{\infty}\leq M$, then, with probability approaching one, we have
\[
\left\Vert \widehat{\pi}_{\mathrm{DNN}}(T,\bs X)-\pi_{0}(T,\bs X)\right\Vert _{P,2}\lesssim n^{-\frac{s_{\pi}}{2s_{\pi}+d+1}}\log^{5}n
\]
and 
\[
\left\Vert \widehat{\bs{\beta}}_{\mathrm{DNNW}}-\bs{\beta}^{*}\right\Vert \lesssim n^{-\frac{s_{\pi}}{2s_{\pi}+d+1}}\log^{5}n\,.
\]
\end{proposition}

Proposition \ref{thm:rate for init DNN beta} gives the convergence rates of $\widehat{\pi}_{\mathrm{DNN}}(T,\bs X)$ and $\widehat{\bs{\beta}}_{\mathrm{DNNW}}$,
whose proof is relegated to Section~\ref{Proof:thm:rate for pi} of the supplementary material. It can be seen from the proof that $\widehat{\pi}_{\mathrm{DNN}}$ has a non-negligible bias, and thus, it is not root-$n$ consistent. Furthermore, the bias of $\widehat{\pi}_{\mathrm{DNN}}$ is carried over to $\widehat{\bs{\beta}}_{\mathrm{DNNW}}$.

\subsection{Asymptotic Distribution of $\widehat{\bs{\beta}}_{\mathrm{BNNW}}$}

In this section, we establish the consistency and asymptotic normality
of our proposed BNNW estimator. Furthermore, we show that the asymptotic
variance of our estimator reaches the semiparametric efficiency bound
(\citealp{bickel1993Efficient,tsiatis2006Semiparametric}). We first state the assumptions.

\begin{assumption}
\label{assu:near zero ass}Recall that $h(Y,T;\bs{\beta})=L^{\prime}(Y,g(T;\boldsymbol{\beta}))\cdot\partial g(T;\bs{\beta})/\partial\bs{\beta}$. 
For any $k=1,\ldots,K$, 
\[
\left\Vert \frac{1}{n}\sum_{i\in I_{k}}\hat{\pi}_{\mathrm{BNN}}^{(k)}(T_{i},\bs X_{i})h(Y_{i},T_{i};\hat{\bs{\beta}}_{\mathrm{DNNW}}^{(-k)})\right\Vert =\inf_{\bs{\beta}\in\varTheta}\left\Vert \frac{1}{n}\sum_{i\in I_{k}}\hat{\pi}_{\mathrm{BNN}}^{(k)}(T_{i},\bs X_{i})h(Y_{i},T_{i};\bs{\beta})\right\Vert +o_{P}(N^{-1/2}).
\]
\end{assumption}
\begin{assumption}
\label{assu:VC type class-mu} Recall that $\mu_{0}(T,\bs X;\bs{\beta})=\e\left[h(Y,T;\bs{\beta})\mid T,\bs X\right]$.
For any $1\leq j\leq p$, let $\mu_{0j}(T,\bs X;\bs{\beta})$ be the
$j$-th component of $\mu_{0}(T,\bs X;\bs{\beta})$. The function class
$\mathcal{F}_{j}:=\left\{ (T,\bs X)\mapsto\mu_{0j}(T,\bs X;\bs{\beta}):\bs{\beta}\in\varTheta\right\} $
attached with measurable envelop $F_{j}:=\sup_{\bs{\beta}\in\varTheta}\left|\mu_{0j}(T,\bs X;\bs{\beta})\right|$
satisfies 
\[
\sup_{Q}\log N\left(\epsilon\left\Vert F_{j}\right\Vert _{Q,2},\mathcal{F}_{j},\left\Vert \cdot\right\Vert _{Q,2}\right)\leq\nu\log\left(a/\epsilon\right)\text{ uniformly over }0<\epsilon\leq1,
\]
where $a,\nu>0$ are constants and $\sup_{Q}$ is taken over all finitely
discrete probability measures. 
\end{assumption}
\begin{assumption}
\label{assu:L2 continuous}For any $1\leq j\leq p$, let $h_{j}(Y,T;\bs{\beta})$
be the $j$-th component of $h(Y,T;\bs{\beta})$ and the function class
$\left\{ h_{j}(Y,T;\bs{\beta}):\bs{\beta}\in\varTheta\right\} $ satisfies
\[
\sup_{\bs{\beta}:\left\Vert \bs{\beta}-\bs{\beta}^{*}\right\Vert <\delta}\e\left[\left|h_{j}(Y,T;\bs{\beta})-h_{j}(Y,T;\bs{\beta}^{*})\right|^{2}\right]^{1/2}\leq a\cdot\delta^{b}
\]
for any small $\delta>0$ and for some finite positive constants $a$
and $b$. 
\end{assumption}

\begin{assumption}[Rate doubly robust]
\label{assu:conver rate for nuisance}Let $\zeta_{\mu},\zeta_{\pi},\zeta_{u},\zeta_{\beta},c^{*}\geq0$
be some finite constants.
\begin{enumerate}[label=(\alph*)]
\item For any $1\leq j\leq p$, let $\mu_{0j}(t,\bs x;\bs{\beta})$ be
the $j$-th component of $\mu_{0}(t,\bs x;\bs{\beta})$. $\mu_{0j}(t,\bs x;\bs{\beta})$
is twice continuously differentiable in $\bs{\beta}\in\varTheta$
and $\e\left[\sup_{\bs{\beta}\in\varTheta}\left\Vert \partial^{2}\mu_{0j}(T,\bs X;\bs{\beta})/(\partial\bs{\beta}\partial\bs{\beta}^{\trans})\right\Vert ^{2}\right]^{1/2}$ $\leq c^{*}$.
\item Either one of the following conditions holds:
\begin{enumerate}[label=(\roman*)]
\item $c^{*}\neq0$, $\zeta_{\pi}+\zeta_{\mu}\geq1/2$, $\zeta_{\pi}+\zeta_{u}+\zeta_{\beta}\geq1/2$,
$\zeta_{\pi}+2\zeta_{\beta}\geq1/2$, $\zeta_{\beta}>0$;
\item $c^{*}=0$, $\zeta_{\pi}+\zeta_{\mu}\geq1/2$, $\zeta_{\pi}+\zeta_{u}+\zeta_{\beta}\geq1/2$,
$\zeta_{\beta}+\zeta_{u}>0$;
\end{enumerate}
\item For a constant $\rho_{0}>0$ and a sequence of non-negative numbers
$\rho_{N}$ satisfying $\rho_{N}=o((\log N)^{-1})$ as $N\to\infty$,
it holds that, for any $1\leq k\le K$, with probability approaching
one,
\begin{align*}
\left\Vert \widehat{\mu}^{(-k)}(T,\bs X;\widehat{\bs{\beta}}_{\mathrm{DNNW}}^{(-k)})-\mu_{0}(T,\bs X;\widehat{\bs{\beta}}_{\mathrm{DNNW}}^{(-k)})\right\Vert _{P,2} & \leq\rho_{N}N^{-\zeta_{\mu}},\\
\left\Vert \widehat{\partial_{\bs{\beta}}\mu}^{(-k)}(T,\bs X;\widehat{\bs{\beta}}_{\mathrm{DNNW}}^{(-k)})-\partial_{\bs{\beta}}\mu_{0}(T,\bs X;\widehat{\bs{\beta}}_{\mathrm{DNNW}}^{(-k)})\right\Vert _{P,2} & \leq\rho_{0}N^{-\zeta_{u}},\\
\left\Vert \widehat{\pi}_{\mathrm{DNN}}^{(-k)}(T,\bs X)-\pi_{0}(T,\bs X)\right\Vert _{P,2}\leq\rho_{N}N^{-\zeta_{\pi}},\ \left\Vert \widehat{\bs{\beta}}_{\mathrm{DNNW}}^{(-k)}-\bs{\beta}^{*}\right\Vert  & \leq\rho_{0}N^{-\zeta_{\beta}}.
\end{align*}
\end{enumerate}
\end{assumption}
\begin{assumption}
\label{assu:moment bound conditions}Let $M>0$ and $q>2$ be some
finite constants. For any $1\leq k\le K$, 
\begin{enumerate}[label=(\alph*)]
\item $\sup_{(t,\bs x)\in\mathcal{T}\times\mathcal{X}}\pi_{0}(t,\bs x)\leq M$;
\item it holds that $\sup_{t\in\mathcal{T},\bs x\in\mathcal{X}}\left|\widehat{\pi}_{\mathrm{DNN}}^{(-k)}(t,\bs x)\right|\leq M$,
$\left\Vert \widehat{\partial_{\bs{\beta}}\mu}^{(-k)}(T,\bs X;\widehat{\bs{\beta}}_{\mathrm{DNNW}}^{(-k)})\right\Vert _{P,(2q)\vee(1/\zeta_{\pi})}\leq M$,
$\left\Vert \widehat{\mu}^{(-k)}(T,\bs X;\widehat{\bs{\beta}}_{\mathrm{DNNW}}^{(-k)})\right\Vert _{P,(2q)\vee(1/\zeta_{\pi})}\leq M$
with probability approaching one, where $\zeta_{\pi}$ is from Assumption
\ref{assu:conver rate for nuisance};
\item 
the minimum singular value of $\e\left[\widehat{\pi}_{\mathrm{DNN}}^{(-k)}(T,\bs X)^{2}\hat{\bs{\xi}}^{(k)}(T,\bs X)\hat{\bs{\xi}}^{(k)}(T,\bs X)^{\trans}\mid(T_{i},\bs X_{i},Y_{i})_{i\notin I_{k}}\right]$
is larger than some constant $c_{\xi}>0$ with probability approaching
one;
\item $\e\left[\sup_{\bs{\beta}\in\varTheta}\left\Vert \partial_{\bs{\beta}}\mu_{0}(T,\bs X;\bs{\beta})\right\Vert ^{2}\right]<\infty$,
$\e\left[\sup_{\bs{\beta}\in\varTheta}\left\Vert h(Y,T;\bs{\beta})\right\Vert ^{2q}\right]<\infty$.
\end{enumerate}
\end{assumption}
Assumption~\ref{assu:near zero ass} allows for an approximate minimizer solution of (\ref{eq:minimization step}), which is useful if the empirical
minimization problem (\ref{eq:minimization step}) has no exact minimizer.
Similar assumptions are also employed by \citet{ai2003Efficient,chernozhukov2018Double,kallus2022Localized,chen2003Estimation}. Similar to Assumption \ref{assu:VC type class}, Assumption \ref{assu:VC type class-mu}
can be verified for a large variety of the loss function $L(y,z)$
and the outcome model $g(t;\bs{\beta})$. Assumption \ref{assu:L2 continuous} only requires $h_{j}(Y,T;\bs{\beta})$
to be $L_{2}$-continuous at $\bs{\beta}^{*}$, it still allows $h_{j}(Y,T;\bs{\beta})$
to be discontinuous so that our theory is applicable to estimating
the QTE. Assumption~\ref{assu:conver rate for nuisance} imposes conditions
on the convergence rate of the nuisance estimators. It can be verified
by Proposition~\ref{thm:rate for init DNN beta} and many other existing results on machine learning convergence rates \citep[e.g.,][]{chi2022Asymptotic,jiao2023Deep}.
This assumption implies that our estimator
is ``rate doubly robust'' in that we require the product of the
convergence rates of $\hat{\pi}_{\mathrm{DNN}}^{(-k)}(\cdot,\cdot)$ and $\hat{\mu}^{(-k)}(\cdot,\cdot;\bs{\beta})$
(or $\widehat{\bs{\beta}}_{\mathrm{DNNW}}^{(-k)}$) to be $o(N^{-1/2})$.\footnote{Our method also enjoys the doubly robustness property in the sense of \cite{bang2005Doubly}, see Section~\ref{sec:doubly robust} in the supplementary material.} For binary ATE,
where $L(y,z)=(y-z)^{2}/2$ and $g(t;\bs{\beta})=\beta_{1}t+\beta_{0}(1-t)$,
it can be shown that $\partial_{\bs{\beta}}\mu_{0}(t,\bs x;\bs{\beta})=\diag\{1-t,t\}$
(see Section~\ref{subsec:The-discussion-of}), giving $c^{*}=0$ in Assumption \ref{assu:conver rate for nuisance}(a) and corresponding to Assumption \ref{assu:conver rate for nuisance}(b)(ii). Then, we can take $\widehat{\partial_{\bs{\beta}}\mu}^{(-k)}(t,\bs x;\bs{\beta})=\diag\{1-t,t\}$, achieving $\zeta_u = \infty$. Thereby,  Assumption~\ref{assu:conver rate for nuisance}
reduces to requiring $\zeta_{\mu},\zeta_{\pi}\geq0$ and $\zeta_{\pi}+\zeta_{\mu}\geq1/2$, which is consistent with the ``rate doubly robust'' result in the
literature \citep{chernozhukov2018Double}, and does not impose any rate requirements on $\zeta_\beta$ of the initial estimator $\bs{\widehat{\beta}}_{\mathrm{DNNW}}^{(-k)}$. Thus, we can simply take $\bs{\widehat{\beta}}_{\mathrm{DNNW}}^{(-k)}=\bs 0$
and use samples in $I_{-k}$ to construct estimator $\widehat{\mu}^{(-k)}(\cdot,\cdot;\bs 0)$
in our BNNW Algorithm \ref{alg:BNNW}. We next consider the estimation of binary QTE.
Provided that we have a consistent estimator $\widehat{\partial_{\bs{\beta}}\mu}^{(-k)}(\cdot,\cdot;\bs{\beta})$
for $\partial_{\bs{\beta}}\mu_{0}(\cdot,\cdot;\bs{\beta})$, Assumption
\ref{assu:conver rate for nuisance} is less restrictive than condition
vii of Theorem 3 in \citet{kallus2022Localized}, which requires that
$\zeta_{\pi}+\zeta_{\mu}\geq1/2$ and $\zeta_{\pi}+\zeta_{\beta}\geq1/2$.
If a consistent estimator for $\partial_{\beta}\mu_{0}(\cdot,\cdot;\bs{\beta})$
is unavailable, Assumption \ref{assu:conver rate for nuisance} reduces
to the same condition as vii of Theorem 3 in \citet{kallus2022Localized}. Assumption \ref{assu:moment bound conditions} imposes some moment conditions concerning $\pi_{0}(\cdot,\cdot)$, $\mu_{0}(\cdot,\cdot;\bs{\beta})$
and $\partial_{\bs{\beta}}\mu_{0}(\cdot,\cdot;\bs{\beta})$ and their
estimates. 
\begin{thm}
\label{thm:asymptotic normal} Recall the definitions of $\Sigma_0$ and $\psi(Y,T,\bs X;\bs{\beta}^{*},\pi_0, \mu_0(T,\bs{X};\bs{\beta}^*))$ in Section~\ref{sec:problem}. Re-denote the latter as  $\psi(Y,T,\bs X;\bs{\beta}^{*})$ for notational simplicity.
Under Assumptions~\ref{assu:unconfounded}, \ref{assu:regularity conditions}-\ref{assu:VC type class},
and \ref{assu:near zero ass}-\ref{assu:moment bound conditions},
the BNNW estimator defined in (\ref{eq:main estimator}) admits 
\[
\hat{\bs{\beta}}_{\mathrm{BNNW}}-\bs{\beta}^{*}=-\frac{1}{N}\sum_{i=1}^{N}\Sigma_{0}^{-1}\psi(Y_{i},T_{i},\bs X_{i};\bs{\beta}^{*})+o_{P}(N^{-1/2}),
\]
and $\sqrt{N}\left(\hat{\bs{\beta}}_{\mathrm{BNNW}}-\bs{\beta}^{*}\right)\tod N(\bs 0,V_{\mathrm{eff}})$,
where $V_{\text{eff}}:=\Sigma_{0}^{-1}\e\left[\psi(Y,T,\bs X;\bs{\beta}^{*})\psi(Y,T,\bs X;\bs{\beta}^{*})^{\trans}\right]\Sigma_{0}^{-1}$
attains the semiparametric efficiency bound, the least possible variance achievable by a regular estimator.
\end{thm}
Theorem \ref{thm:asymptotic normal} implies that our BNNW
algorithm yields a semiparametric efficient estimator for $\bs{\beta}^{*}$
under relatively mild conditions. It is worth noting that if the stabilized
weighting function $\pi_0$ is known, one can replace $\widehat{\pi}$ in the minimization \eqref{eq:naive minimization step} by $\pi_0$ to obtain an estimator of $\bs{\beta}^*$. Note that the estimator proposed by \citet[p.786]{ai2021Unified} is a special case of \eqref{eq:naive minimization step}. Using their argument, it can be shown that the resulting estimator is inefficient. On the contrary,
Theorem~\ref{thm:asymptotic normal} shows that the BNNW
estimator remains semiparametric efficient even when replacing the initial weighting $\widehat{\pi}_{\text{DNN}}$ by the true
$\pi_0$. Therefore, calibrating the stabilized weighting function is not just for reducing bias, but also for increasing the robustness of the resulting estimator.

\section{Inference \label{sec:Inference}}

In this section, we employ the weighted bootstrap method to conduct
the inference for $\bs{\beta}^{*}$. 

\begin{myalgorithm}[Weighted Bootstrap]Let $B>1$ be a constant.
The weighted bootstrap procedure has the following steps: for $b=1,2,\ldots,B$, 
\begin{enumerate}
\item Generate independent and identically distributed non-negative sub-exponential random variables $U_{1},\ldots,U_{N}$,
independent of $\{(T_{i},\bs X_{i},Y_{i})\}_{i=1}^{N}$,
with $\e(U_{1})=\var(U_{1})=1$; 
\item For $k=1,2,\ldots,K$, 
\begin{enumerate}
\item Calculate
\[
\widehat{w}^{(k)}_{i}=\rho^{\prime}\left\{U_i\widehat{\pi}^{(-k)}_{\mathrm{DNN}}(T_{i},\bs X_{i})\hat{\bs{\lambda}}_k^{\trans}\widehat{\bs{\xi}}^{(k)}(T_{i},\bs X_{i})\right\}\ \text{for}\ i\in \{1,...,n\},
\]
where $\hat{\bs{\lambda}}_k$ is the unique maximizer of the refined concave function 
{\small\[
\hat{G}(\bs{\lambda}):=\frac{1}{n}\sum_{i\in I_k}\rho\left\{U_i\widehat{\pi}^{(-k)}_{\mathrm{DNN}}(T_{i},\bs X_{i})\bs{\lambda}^{\trans}\widehat{\bs{\xi}}^{(k)}(T_{i},\bs X_{i})\right\}-\frac{1}{n(n-1)}\sum_{j\neq l \in I_{k}}U_jU_l\bs{\lambda}^{\trans}\widehat{\bs{\xi}}^{(k)}(T_{j},\bs X_{l})\,,
\]}
where the nuisances estimates are taken from Step 1 in Algorithm~\ref{alg:BNNW}; 
\item solve $\hat{\bs{\beta}}_{b}^{(k)}$ by the minimization problem: 
$$
\hat{\bs{\beta}}_{b}^{(k)}=\underset{\bs{\beta}\in\R^{p}}{\arg\min}\sum_{i\in I_{k}}\widehat{w}_{i}^{(k)}U_{i}\widehat{\pi}_{\mathrm{DNN}}^{(-k)}(T_{i},\bs X_{i})L(Y_{i},g(T_{i};\bs{\beta}))\,;$$
\end{enumerate}
\item Define the weighted bootstrap estimator as $\hat{\bs{\beta}}_{b}=K^{-1}\sum_{k=1}^{K}\hat{\bs{\beta}}_{b}^{(k)}$.
\end{enumerate}
\end{myalgorithm}

In our simulation studies and real applications, we take $U_i\sim 2\cdot\mathrm{Bernoulli}(0.5) $ for $1\leq i \leq N$. Our weighted bootstrap method does not require estimating the nuisances
for every $1\leq b\leq B$, but uses the estimates from our BNNW Algorithm~\ref{alg:BNNW}.
Therefore, our inference procedure is time-efficient. Before validating our weighted bootstrap procedure, we need an assumption, which is
similar to Assumption~\ref{assu:near zero ass}. 
\begin{assumption}
\label{assu:bootstra near zero}For any $b=1,2,\ldots,B$ and $k=1,2,\ldots,K$,
{\footnotesize\[
\left\Vert \frac{1}{n}\sum_{i\in I_{k}}\widehat{w}_{i}^{(k)}U_{i}\widehat{\pi}_{\mathrm{DNN}}^{(-k)}(T_{i},\bs X_{i})h(Y_{i},T_{i};\hat{\bs{\beta}}_{b}^{(k)})\right\Vert =\inf_{\bs{\beta}\in\varTheta}\left\Vert \frac{1}{n}\sum_{i\in I_{k}}\widehat{w}_{i}^{(k)}U_{i}\widehat{\pi}_{\mathrm{DNN}}^{(-k)}(T_{i},\bs X_{i})h(Y_{i},T_{i};\bs{\beta})\right\Vert +o_{P}(N^{-1/2})\,.
\]}
\end{assumption}
\begin{thm}
\label{thm:MB theorem}Under Assumptions \ref{assu:unconfounded},
\ref{assu:regularity conditions}-\ref{assu:VC type class}, and \ref{assu:VC type class-mu}-\ref{assu:bootstra near zero},
for any $b=1,2,\ldots,B$, 
\[
\hat{\bs{\beta}}_{b}-\bs{\beta}^{*}=-\frac{1}{N}\sum_{i=1}^{N}U_{i}\Sigma_{0}^{-1}\psi(Y_{i},T_{i},\bs X_{i};\bs{\beta}^{*})+o_{P}(N^{-1/2}).
\]
This, combined with Theorem \ref{thm:asymptotic normal} yields that
\[
\hat{\bs{\beta}}_{b}-\hat{\bs{\beta}}_{\mathrm{BNNW}}=-\frac{1}{N}\sum_{i=1}^{N}\left(U_{i}-1\right)\Sigma_{0}^{-1}\psi(Y_{i},T_{i},\bs X_{i};\bs{\beta}^{*})+o_{P}(N^{-1/2}).
\]
\end{thm}
Theorem \ref{thm:MB theorem} suggests that for any vector $\bs{\nu}\in\R^{p}$,
the $(1-\alpha)$ bootstrapped equitailed confidence interval for
$\bs{\nu}^{\trans}\bs{\beta}^{*}$ can be constructed as 
$
\left[\widehat{c}_{\alpha/2},\widehat{c}_{1-\alpha/2}\right]$, where $\widehat{c}_{\alpha/2}$ and $\widehat{c}_{1-\alpha/2}$ are the empirical quantiles of $\{ \bs{\nu}^{\trans}\hat{\bs{\beta}}_{b}:1\leq b\leq B\} $ at level $\alpha/2$ 
 and $1-\alpha/2$, respectively.
\begin{rem}
Section \ref{sec:Theoretical-Properties} establishes the asymptotic
normality of our BNNW estimator under mild conditions. This
suggests that a valid confidence
interval can also be constructed given a consistent estimate of the asymptotic variance. A direct way to estimate the asymptotic
variance $V$ is based on the influence function ${\Sigma}_{0}^{-1}\psi(Y,T,\bs X;\bs{\beta}^{*})$.
Specifically, one can estimate $V$ by 
\[
\hat{V}=\frac{1}{N}\sum_{i=1}^{N}\widehat{\Sigma}_{0}^{-1}\hat{\psi}(Y_{i},T_{i},\bs X_{i};\hat{\bs{\beta}})\hat{\psi}(Y_{i},T_{i},\bs X_{i};\hat{\bs{\beta}})^{\trans}\widehat{\Sigma}_{0}^{-1},
\]
where $\hat{\psi}(Y,T,\bs X;\hat{\bs{\beta}})$ and $\widehat{\Sigma}_{0}^{-1}$ are the estimates of
$\psi(Y,T,\bs X;\bs{\beta}^{*})$ and $\Sigma_{0}^{-1}$, respectively. This method, however, generally
yields poorly-performed estimator since estimating $\Sigma_{0}^{-1}\psi(Y,T,\bs X;\bs{\beta}^{*})$
requires additional estimates for $\e\left[\mu_{0}(T,\bs X;\bs{\beta}^{*})\pi_{0}(T,\bs X)\mid T\right]$,
$\e\left[\mu_{0}(T,\bs X;\bs{\beta}^{*})\pi_{0}(T,\bs X)\mid\bs X\right]$
and $\Sigma_{0}$, which is more challenging compared to the bootstrap-based
method, especially when the dimension of the covariates $\bs X$ is
large. 
\end{rem}

\section{Monte Carlo Studies \label{sec:Monte-Carlo-studies}}
To assess the performance of the BNNW estimator, we conduct Monte Carlo simulations under various scenarios. In Section \ref{sec:SimBinary}, we investigate the estimator’s performance with a binary treatment, followed by an analysis of its behavior with a continuous treatment in Section \ref{sec:SimContinuous}. 

\subsection{Scenario 1: Binary Treatment \label{sec:SimBinary}}


For binary treatment, we consider a setting similar to that in \citet{donald2014}.
Let $U_{xj},U_{t},U_{y_{0}},U_{y_{1}}$ ($1\leq j\leq d$) be
independent uniform random variables over $[0,1]$, and $\bs X=(X_{1},\ldots,X_{d})^{\trans}$
be a vector of covariates with dimension $d$, $\bs{\gamma}_{b}\in\R^{d}$
be a predetermined constant that ensures $\bs X^{\trans}\bs{\gamma}_{b}\in(0,1)$.
The data-generating process with a binary treatment (DGP-B) is as follows:
\\
\textbf{DPG-B} : 
\[
\begin{aligned} & X_{j}=0.3+0.4U_{xj},\ j=1,\ldots,d\ \ \text{and}\ \ T=1(U_{t}<\bs X^{\trans}\bs{\gamma}_{b});\\
	& Y(0)=1(U_{y_{0}}\le\bs X^{\trans}\bs{\gamma}_{b})\frac{U_{y_{0}}^{2}}{\bs X^{\trans}\bs{\gamma}_{b}}+1(U_{y_{0}}>\bs X^{\trans}\bs{\gamma}_{b})U_{y_{0}};\\
	& Y(1)=1(U_{y_{1}}\le1-\bs X^{\trans}\bs{\gamma}_{b})\frac{2U_{y_{1}}^{2}}{1-\bs X^{\trans}\bs{\gamma}_{b}}+1(U_{y_{1}}>1-\bs X^{\trans}\bs{\gamma}_{b})2U_{y_{1}}.
\end{aligned}
\]


To evaluate the performance of our method under setting with a binary
treatment, we estimate the ATE and QTE over samples simulated by DPG-B, with dimension of covariates $d=50$, sample size $N$ varies from 300 to 5000, and different quantiles. For each combination of configurations, we generate $S=100$ replications. We assess the performance of the point estimation by computing the bias, standard error (SE), and root mean squared error
(RMSE) of the $S=100$ estimates; see Section~\ref{sec:Num_details} of the supplementary materials for the formal definitions of these metrics.
For interval estimation,
the coverage probability (CP) and average width (AW) of 95\% confidence intervals are calculated 
over the $S=100$ replications.
We compare the ATE and QTE estimation results between our BNNW method and alternative ones.

\subsubsection{ATE}
We first compare our method to a double machine learning method in the literature \citep{chernozhukov2018Double}, which leverages a cross-fitted augmented inverse probability weighting (AIPW) as defined below, for ATE estimation. 
$$
{\small \hat{\tau}_{\text{AIPW}}  =\frac{1}{N}\sum_{k=1}^{K}\sum_{i\in I_{k}}\left(\hat{m}_{1}^{(-k)}(\boldsymbol{X}_{i})-\hat{m}_{0}^{(-k)}(\boldsymbol{X}_{i})+T_{i}\frac{Y_{i}-\hat{m}_{1}^{(-k)}(\boldsymbol{X}_{i})}{\hat{e}^{(-k)}(\boldsymbol{X}_{i})}-(1-T_{i})\frac{Y_{i}-\hat{m}_{0}^{(-k)}(\boldsymbol{X}_{i})}{1-\hat{e}^{(-k)}(\boldsymbol{X}_{i})}\right)\,,}
$$

where $\hat{e}^{(-k)}(\boldsymbol{X})$ and $\hat{m}_{t}^{(-k)}(\boldsymbol{X})$ are some estimates of the propensity score $P(T=1\lvert\boldsymbol{X})$ and the outcome regression $\e(Y|\bs{X}, T=t)$, respectively, 
using the sample in $I_{-k}$.
For both BNNW and AIPW methods, we split the sample into $K=5$ folds for cross-fitting. We compute $\hat{e}^{(-k)}(\boldsymbol{X})$ and $\hat{m}_{t}^{(-k)}(\boldsymbol{X})$ for AIPW and estimate $\bs\xi(T,\bs X)$ for BNNW using LightGBM with the number of boosted
trees as $300$ and maximum tree leaves as 10. To obtain $95\%$ confidence
intervals, the weighted bootstrap is conducted with $B=599$.

\begin{table}[!htbp]
  \caption{Bias, SE, RMSE of estimated ATE and coverage probability (CP), average width (AW) of the respective $95\%$ confidence interval, using BNNW or AIPW estimators over $100$ replications of simulated dataset with a binary treatment across different sample sizes.}
  \label{tab:sim-binary-ate}
  \centering
  \begin{tabular}{@{} c *{5}{c} *{5}{c} @{}}
    \toprule
    & \multicolumn{5}{c}{AIPW} & \multicolumn{5}{c}{BNNW} \\
    \cmidrule(lr){2-6} \cmidrule(lr){7-11}
    N & Bias & SE & RMSE & CP & AW & Bias & SE & RMSE & CP & AW \\
    \midrule
    300  & 0.0781 & 0.070 & 0.0849 & 0.69 & 0.235 & 0.0022 & 0.061 & 0.0512 & 0.98 & 0.258 \\
    500  & 0.0670 & 0.048 & 0.0698 & 0.71 & 0.187 & 0.0040 & 0.052 & 0.0406 & 0.96 & 0.202 \\
    1000 & 0.0349 & 0.036 & 0.0419 & 0.82 & 0.140 & 0.0094 & 0.033 & 0.0278 & 0.95 & 0.144 \\
    5000 & 0.0097 & 0.017 & 0.0157 & 0.91 & 0.065 & 0.0049 & 0.017 & 0.0144 & 0.94 & 0.065 \\
    \bottomrule
  \end{tabular}
\end{table}

The evaluation metrics are reported in Table \ref{tab:sim-binary-ate}, from which it is clear that BNNW estimator outperforms the AIPW estimator for both the point and interval estimation, especially when the sample size is small. The standard errors of the BNNW and AIPW estimators behave similarly, while the bias of BNNW estimator is much smaller, suggesting that our BNNW estimator effectively reduces bias.

\begin{figure}[!htbp]
	\begin{centering}
		\includegraphics[width=0.45\textwidth]{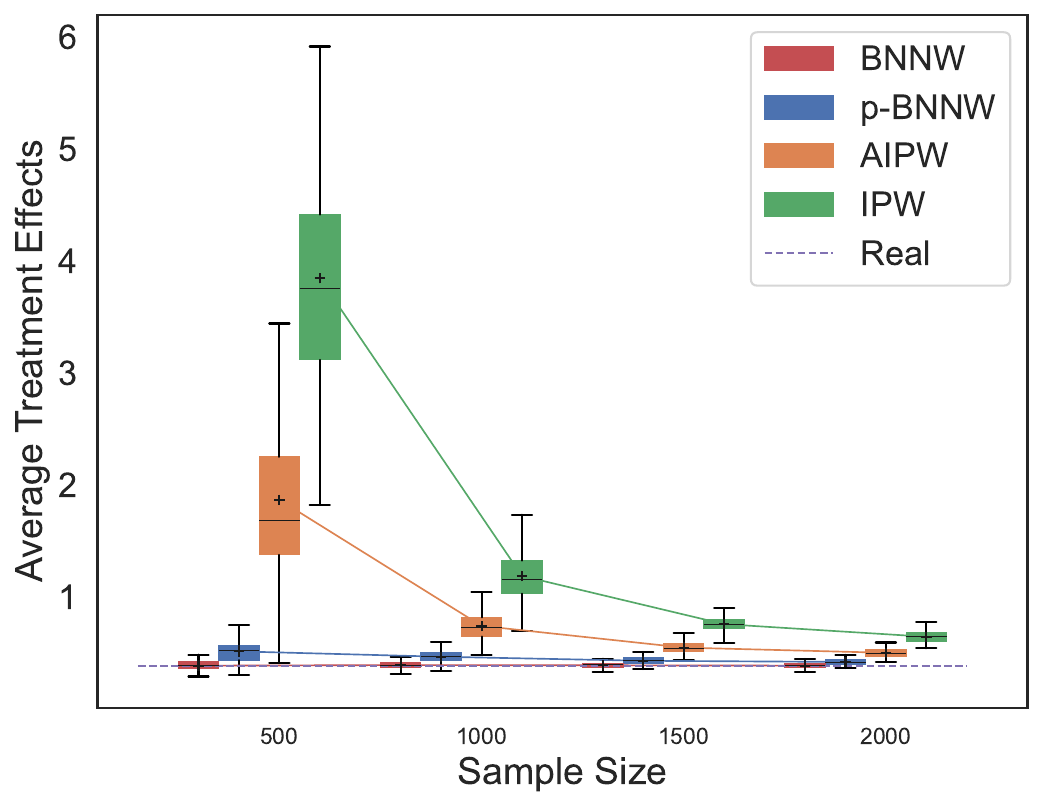}
        \hspace{0.1cm}
        \includegraphics[width=0.46\textwidth]{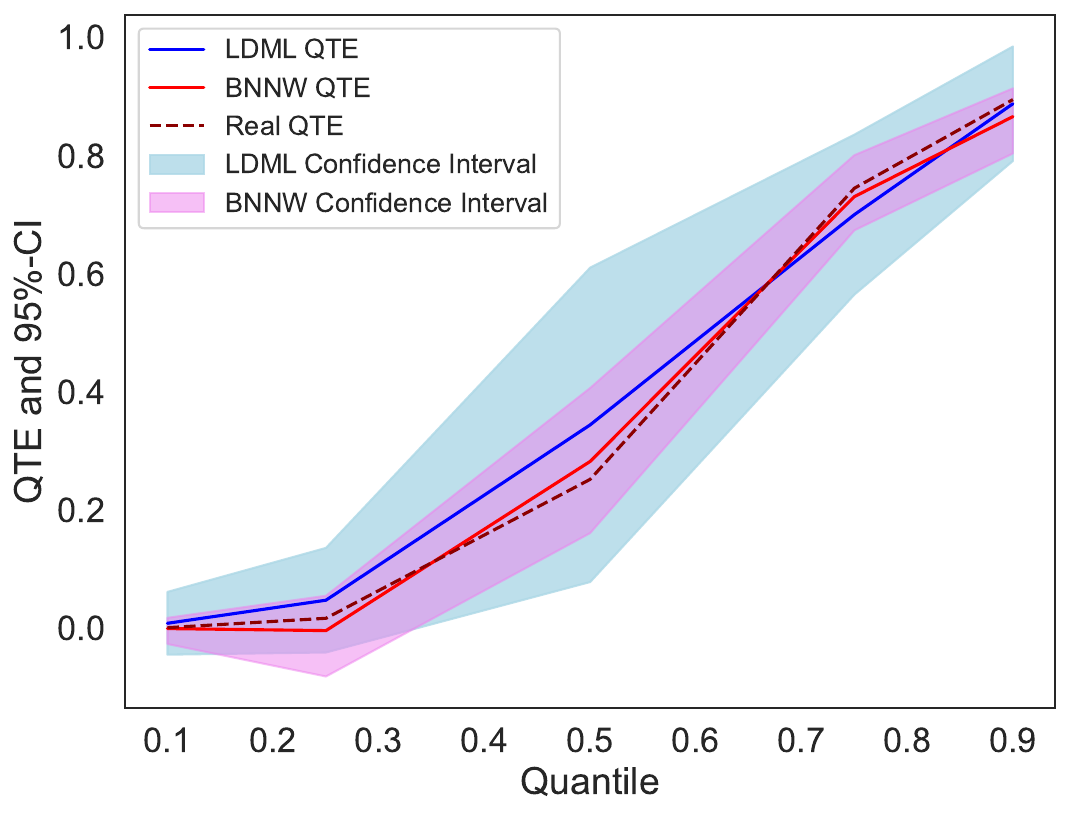}
		
		\par\end{centering}
	\caption{Simulation results under a binary treatment setting. Left panel: Box plots of the estimated ATE using different methods over $100$ replications of each sample size setting with $d=50$. Right panel: Estimated QTE and the respective $95\%$ confidence interval across different quantile levels on a simulated data set with $N=2000$ and $d=50$.}
	\label{fig:sim-ate-qte}
\end{figure}

To better demonstrate how the BNNW estimator improves the finite sample estimation of ATE, we further involve the non-debiased inverse probability weighting (IPW) estimator, and p-BNNW estimator that corrects the bias of IPW using our covariate balancing weights, in the comparison (see the left panel of Figure~\ref{fig:sim-ate-qte}). More specificaly, these estimators are defined as follows: 
\begin{align*}
\hat{\tau}_{\text{IPW}} & =\frac{1}{N}\sum_{k=1}^{K}\sum_{i\in I_{k}}\left(\frac{T_{i}Y_{i}}{\hat{e}^{(-k)}(\boldsymbol{X}_{i})}-\frac{(1-T_{i})Y_{i}}{1-\hat{e}^{(-k)}(\boldsymbol{X}_{i})}\right)\,,\\
\hat{\tau}_{\text{p-BNNW}} & =\frac{1}{N}\sum_{k=1}^{K}\sum_{i\in I_{k}}\left(\frac{T_{i}\hat{\omega}_{i}^{(k)}Y_{i}}{\hat{e}^{(-k)}(\boldsymbol{X}_{i})}-\frac{(1-T_{i})\hat{\omega}_{i}^{(k)}Y_{i}}{1-\hat{e}^{(-k)}(\boldsymbol{X}_{i})}\right)\,,
\end{align*}
with $K=5$ and $(\hat{\omega}_{i}^{(k)})_{i\in I_{k}}$
are the calibration weights solving 
\[
\begin{cases}
\min_{\omega_{i}:i\in I_{k}}\ \sum_{i\in I_{k}}D(\omega_{i})\ \text{s.t.}\\
\sum_{i\in I_{k}}\omega_{i}\frac{T_{i}\hat{m}_{1}^{(-k)}(\boldsymbol{X}_{i})}{\hat{e}^{(-k)}(\boldsymbol{X}_{i})}=\sum_{i\in I_{k}}\hat{m}_{1}^{(-k)}(\boldsymbol{X}_{i})\\
\sum_{i\in I_{k}}\omega_{i}\frac{(1-T_{i})\hat{m}_{0}^{(-k)}(\boldsymbol{X}_{i})}{1-\hat{e}^{(-k)}(\boldsymbol{X}_{i})}=\sum_{i\in I_{k}}\hat{m}_{0}^{(-k)}(\boldsymbol{X}_{i})\,,
\end{cases}
\]
where $\hat{e}^{(-k)}(\boldsymbol{X})$ and $\hat{m}_{t}^{(-k)}(\boldsymbol{X})$ are the same as those in $\hat{\tau}_{\text{AIPW}}$.

Note that the AIPW, IPW and p-BNNW estimators all rely on the inverse of estimated propensity scores. While AIPW reduces the bias of IPW via EIF, p-BNNW does so through our proposed balancing approach. As shown in the left panel of Figure~\ref{fig:sim-ate-qte}, p-BNNW improves both the bias and the variance of IPW more effectively than AIPW, highlighting the advantage of our approach. Moreover, computing the balancing weights $(\hat{\omega}_i^{(k)})_{i\in I_k}$ incurs a negligible additional cost compared to estimating the nuisances for the p-BNNW estimator. The slight performance gain of the BNNW estimator over p-BNNW arises from estimating the stabilized weights $\pi_{0}(t,\boldsymbol{x})$ directly, rather than using the inverse of estimated propensity scores. This approach also generalizes to QTE estimation scenarios, continuous treatments, and beyond. 

\subsubsection{QTE}
We compare the performance of BNNW to the localized double machine learning
(LDML) estimator proposed by \citet{kallus2022Localized} for QTE estimation. We again use $K=5$ for cross-fitting and $B=599$ for bootstrapping for both methods. The calibration function $\bs{\xi}(t,\bs{x})$ for BNNW and the nuisance functions for LDML are all estimated using LightGBM under identical settings as those for the ATE case. The visualization of the comparison using one simulated data set is depicted in the right panel of Figure \ref{fig:sim-ate-qte} and the full simulation results are provided in Table \ref{tab:sim-binary-qte} in Section~\ref{sec:Num_details} of the supplementary material.

The simulation results for QTE estimation show that BNNW outperforms LDML in point and interval estimation, achieving lower RMSE and narrower 95\% confidence intervals with comparable coverage. The advantage of BNNW is especially pronounced with smaller sample sizes.

\subsection{Scenario 2: Continuous Treatment \label{sec:SimContinuous}}

To study continuous treatment effects, we use the semi-synthetic version of the Infant Health and Development Program (IHDP) dataset from \citet{hillBayesianNonparametricModeling2011}. The original dataset includes binary treatments with $N=747$ observations and 25 covariates. We adapt the data-generating process of the semi-synthetic IHDP dataset in \citet{gaoVariationalFrameworkEstimating2023} to create continuous treatment $T$ and potential outcome $Y(t)$ based on the covariates $\bs X=(X_1,\ldots, X_{25})^{\trans}$: 
\begin{align*}
 & T=\{1+\exp(\widetilde{T})\}^{-1}\,, \quad Y^{*}(t)=h(t,\bs{X})+\epsilon_{2}\,,\\
 &\text{where} \quad \widetilde{T}=\frac{3X_{1}}{1+X_{2}}+\frac{3\max\{X_{3},X_{4},X_{5}\}}{0.2+\min\{X_{3},X_{4},X_{5}\}}+3\tanh\left(\frac{5\sum_{j\in\mathcal{J}_{1}}X_{j}}{\lvert\mathcal{J}_{1}\lvert}\right)-6+\epsilon_{1}\,,\\
& h(t,\bs{X})=(-0.8+3.2t-3.2t^{2})\cdot\left\{\tanh\left(\frac{5\sum_{j\in\mathcal{J}_{2}}X_{j}}{\lvert\mathcal{J}_{2}\lvert}\right)+3\exp\left(\frac{0.2(X_{1}-X_{5})}{0.1+\min\{X_{2},X_{3},X_{4}\}}\right)\right\}\,,
\end{align*}
$\mathcal{J}_1=\{3,6,7,8,9,10,11,12,13,14\}$, $\mathcal{J}_2=\{15,16,17,18,19,20,21,22,23,24\}$, and $\epsilon_1, \epsilon_2$ are two independent standard normal random variables.

\begin{table}[!htbp]
\caption{The ABias, ASE and ARMSE of the estimated average dosage response function (ADRF) and quantile dosage response function (QDRF) for various quantiles $\tau$ over 100 replications of the semi-synthetic IHDP dataset with a continuous treatment.}
\label{tab-IHDP}
\centering
\begin{tabular}{@{}llc*{5}{c}@{}} 
\toprule
\multirow{3}{*}{ } & \multirow{3}{*}{ } & \multirow{3}{*}{ADRF} & \multicolumn{5}{c}{QDRF} \\
\cmidrule(lr){4-8} 
 & &  & $\tau=0.1$ & $\tau=0.25$ & $\tau=0.5$ & $\tau=0.75$ & $\tau=0.9$\\
\midrule
     & ABias & 0.215 & 0.391 & 0.198 & 0.203 & 0.198 & 0.212 \\
BNNW & ASE   & 0.103 & 0.329 & 0.153 & 0.109 & 0.105 & 0.124 \\
     & ARMSE & 0.247 & 0.524 & 0.256 & 0.236 & 0.231 & 0.255 \\
\addlinespace[0.5em]
     & ABias & 0.619 & 1.173 & 0.516 & 0.326 & 0.254 & 0.226 \\
DNNW & ASE   & 0.078 & 0.251 & 0.131 & 0.098 & 0.094 & 0.119 \\
     & ARMSE & 0.628 & 1.211 & 0.539 & 0.347 & 0.278 & 0.266 \\
\addlinespace[0.5em]
     & ABias & 0.605 & 1.405 & 0.643 & 0.295 & 0.253 & 0.206 \\
GOE  & ASE   & 0.398 & 0.924 & 0.591 & 0.500 & 0.545 & 0.725 \\
     & ARMSE & 0.736 & 1.703 & 0.885 & 0.586 & 0.605 & 0.754 \\
\bottomrule
\end{tabular}
\end{table}

To demonstrate the effectiveness of our proposed estimators, we performed a comparative analysis of the BNNW estimator, the initial DNNW estimator in our method, and the generalized optimization estimator (GOE) from \citet{ai2021Unified}. All methods use the correctly specified model for $g(t,\bs{\beta})$, namely a degree-2 polynomial.  
The evaluation is based on $S=100$ replications of the IHDP-continuous dataset. We evaluated performance across the full range of dosage levels using three metrics: average absolute bias (ABias), average stochastic error (ASE) and average root mean squared error (ARMSE). See Section~\ref{sec:Num_details} of the supplementary materials for formal definitions.

The results are presented in Table~\ref{tab-IHDP}. As expected, DNNW outperforms GOE, particularly in terms of ASE, probably due to the greater flexibility of DNNs in dealing with high-dimensional data. Our BNNW estimator further improves upon DNNW, especially in reducing bias, highlighting the effectiveness of our proposed debiasing method.

\section{Empirical Study \label{sec:Real-data}}

\subsection{Binary Treatment Case: 401(k) Plans Dataset}

To evaluate the performance of the BNNW method in estimating treatment effects with real-world binary treatment data, we conduct an empirical study using the 401(k) dataset from \citet{chernozhukov2004}. The 401(k) plan is a key component of retirement planning for millions of American workers, allowing employees to contribute a portion of their salary on a pre-tax basis, thereby offering a tax-advantaged way to save for retirement.
The publicly available dataset includes 9,915 individuals and 12 covariates related to their demographics and financial status. It has been widely used in previous studies, including \citet{Poterba1994401klans, Poterba1995Do401k, Poterba1996Personal}, which investigated the impact of 401(k) eligibility on individual wealth, and \citet{kallus2022Localized}, which estimated and conducted inference on the QTEs of 401(k) eligibility on net assets.\footnote{They use 401(k) eligibility—not participation—as the treatment variable because `eligibility' satisfies the unconfoundedness assumption when conditioned on key covariates (e.g., age, income, education), as argued in \citet{chernozhukov2004} and supported by the employer-determined nature of eligibility rules \citep{Poterba1994401klans, Poterba1995Do401k, Poterba1996Personal}. In contrast, `participation' is subject to self-selection bias from unobserved factors like savings motivation and financial literacy.}

Using 401(k) eligibility as the binary treatment and net financial assets as the outcome, we reanalyze the impact of 401(k) plans with the BNNW method, conditioning on 9 key covariates. The stabilized weights $\pi_{0}(t,\bs x)$ are estimated via our BNNW, and $\mu_{0}(t,\bs x;\bs{\beta})$ via LightGBM, using $K=5$ cross-fitting folds. Estimates of the ATE
and the $10\%$ to $90\%$ QTEs are reported in
Table~\ref{tab-401k}. Compared to LDML method, the BNNW estimator achieves precision improvements over LDML, notably reducing confidence interval widths by 12.1\% at $\tau=0.1$ and 17.6\% at $\tau=0.5$ for example. This enhancement can translate to more reliable real-world decision-making, where narrower confidence intervals allow statistically robust conclusions to be drawn efficiently with reduced data collection costs.


\begin{table}[htbp]
\centering
\caption{The ATE and QTE estimated by BNNW method and LDML method at different quantile levels, and the respective $95\%$ confidence interval, for the treatment effects of 401(k) eligibility on individual wealth.}
\label{tab-401k}
\begin{tabular}{cc *{3}{S[table-format=5.2]} *{3}{S[table-format=5.2]}}
\toprule
\multirow{3}{*}{} & \multirow{3}{*}{$\tau$} & \multicolumn{3}{c}{BNNW} & \multicolumn{3}{c}{LDML} \\
\cmidrule(lr){3-5} \cmidrule(lr){6-8}
 & & {Estimation} & {95\% Lower} & {95\% Upper} & {Estimation} & {95\% Lower} & {95\% Upper} \\
\midrule
 & 0.1 & 1449.92 & 730.21 & 2314.52 & 1465.00 & 563.77 & 2366.23 \\
 & 0.25 & 982.56 & 605.19 & 1470.33 & 1000.00 & 409.97 & 1590.03 \\
QTE & 0.5 & 4916.32 & 3789.72 & 5607.62 & 4500.00 & 3531.16 & 5468.84 \\
 & 0.75 & 14919.74 & 12443.37 & 16841.91 & 12551.00 & 9803.89 & 15298.11 \\
 & 0.9 & 26189.76 & 20439.22 & 32946.29 & 20707.00 & 14839.39 & 26574.61 \\
\addlinespace
ATE & {--} & 10482.17 & 8206.15 & 13753.91 & {--} & {--} & {--} \\
\bottomrule
\end{tabular}
\end{table}

Our findings align with those of \citet{kallus2022Localized}, confirming that 401(k) eligibility has a positive effect on individual wealth. The effect exhibits a nonlinear pattern -- initially decreasing and then increasing across quantiles -- suggesting that the benefits of eligibility are more substantial for wealthier individuals, while still providing meaningful gains for those with lower financial resources.

\subsection{Continuous Treatment Case: MSF Dataset\label{sec:real data continuous}}

In our real data application with continuous treatment effects, we employ the Mother's Significant Features (MSF) dataset from \citet{msf_data}. 
This dataset encompasses 450 case studies from the Mumbai region, collating comprehensive maternal, paternal, and health-related information gathered through post-childbirth interviews conducted between February 2018 and March 2021. It comprises 130 variables, spanning across various domains such as physical attributes, social factors, lifestyle choices, stress levels, and health outcomes, thus providing a detailed view of complications and outcomes associated with child health, maternal well-being, and the course of pregnancy.

\begin{figure}[!htbp]
	\begin{centering}
		\includegraphics[width=\textwidth]{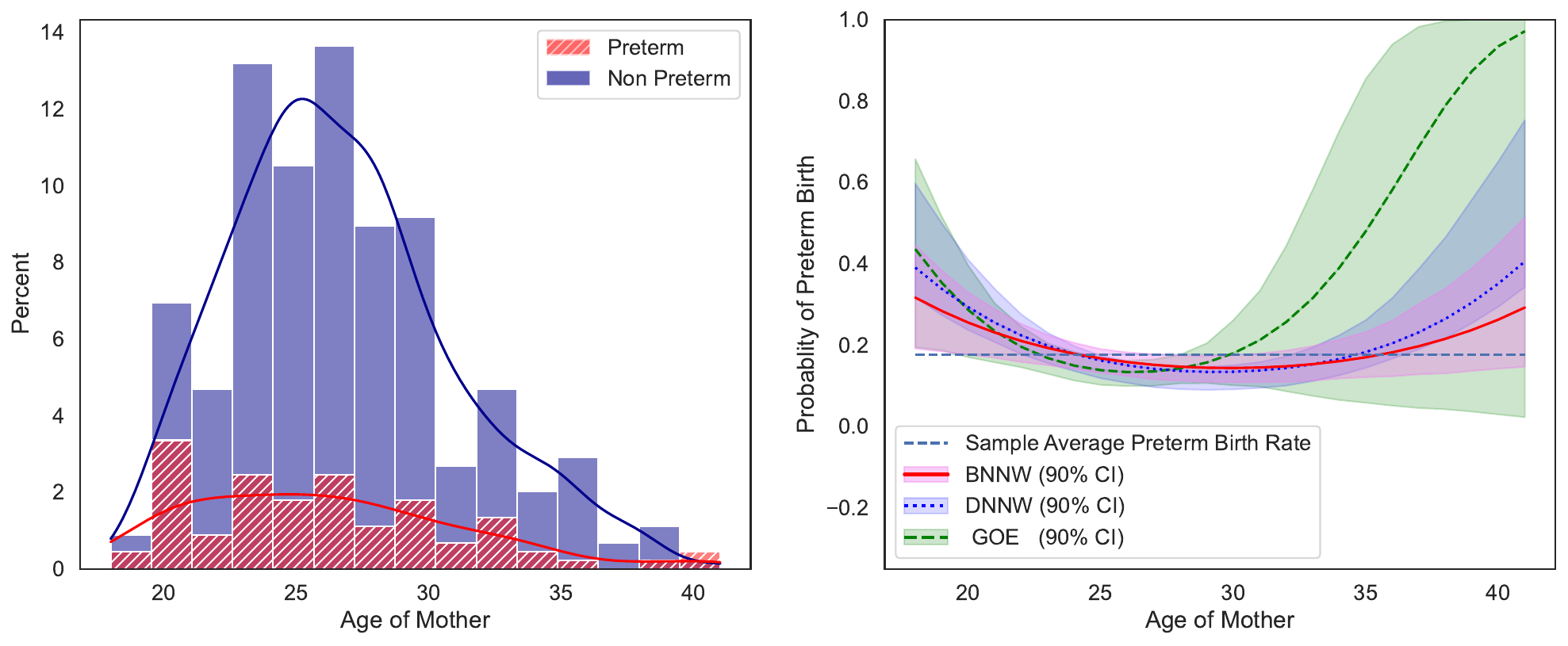}
		
		\end{centering}
        
	\caption{The influence of maternal age on preterm birth rates within the MSF Dataset. Left panel: Histograms of maternal age for both the preterm and non-preterm birth groups. Right panel: Estimated probability of preterm birth and the respective $90\%$ confidence interval.}
	\label{fig:realdata-msf}
\end{figure}

Our investigation is centered on assessing the influence of maternal age, a significant risk factor, on the probability of preterm births. The binary outcome variable indicates preterm birth occurrence (gestational age < 37 weeks). We adopt a quadratic parametric probit model (the model selection will be discussed later) to characterize the likelihood of preterm birth occurrences, defined as $g(t;\bs{\beta})=\Phi\left(\beta_0+\beta_1 t+\beta_2 t^2\right)$, where $\Phi(\cdot)$ is the cumulative distribution function of the standard normal distribution. The associated loss function is formulated as the negative log-likelihood:
$
L(y,g(t;\bs{\beta}))=-y\log g(t;\bs{\beta})-(1-y)\log(1-g(t;\bs{\beta}))
$. Adopting this framework, we implement our BNNW algorithm to estimate the average probability of the treatment dose response. The resultant predicted probability function, along with its $90\%$ confidence intervals, is presented in Figure \ref{fig:realdata-msf}.

To gain an initial insight of the potential impact of maternal age on preterm birth, we first construct comparative histograms of maternal age distributions stratified by preterm birth status in the left panel of Figure \ref{fig:realdata-msf}. Overall, the distribution of maternal age in preterm group is more right-skewed than that in non-preterm group with a heavier right tail, suggesting that younger and older mothers tend to have a higher risk of spontaneous preterm birth. Additionally, the observed proportion of preterm birth among mothers aged 35 and above (23.5\%) is greater than that within the 25 to 35 age range (15.2\%). These findings are in close agreement with the results obtained from the BNNW estimator, as presented in the right panel of Figure \ref{fig:realdata-msf}.

For parametric model selection, we evaluate various polynomial degrees by  comparing Bayesian Information Criterion (BIC) values. 
Among competing models, the quadratic parametric model achieves the lowest BIC value of $421.55$, outperforming the cubic model ($\text{BIC} = 448.31$) and the linear model ($\text{BIC} = 427.16$). Consequently, we select the quadratic model for its optimal balance between goodness-of-fit and complexity. The estimated parameters for the selected model are $\hat{\bs{\beta}}=\left( -1.020 ,\ -0.118,\  0.081\ \right)$, with the respective 95\% confidence intervals: $\beta_0\in( -1.179,\ -0.860),\ \beta_1\in(-0.257,\ 0.019),\ \beta_2\in(0.018,\  0.183).$ 

Compared to DNNW and GOE estimators, BNNW estimator offers a superior balance of flexibility and stability. It outperforms DNNW in handling boundary effects, and provides more stable and reasonable estimates with narrower confidence bands than GOE; see the right panel of Figure~\ref{fig:realdata-msf}.

The estimated counterfactual preterm birth probability curve suggests that women under 25 and over 35 face elevated risks, with probabilities exceeding the overall sample proportion of 17.7\%. In particular, the entire 90\% confidence interval for women under 20 lies above the sample average, indicating a significantly higher risk, whereas the interval for women aged 35 to 41 does not surpass the overall rate significantly. These statistical revelations are consistent with the comprehensive scholarly research within the healthcare domain. They provide critical insights for women contemplating family planning, aiding in the determination of the most favorable timing for conception. Moreover, these findings emphasize the necessity for healthcare facilities to prioritize and intensify support for expectant mothers in these age groups.





\bibliographystyle{dcu}
\bibliography{r}

\end{document}